# SLA Management in Intent-Driven Service Management Systems: A Taxonomy and Future Directions

YOGESH SHARMA*, University of Regina, Canada
DEVAL BHAMARE, Samsung Research (SRUK), United Kingdom
NISHANTH SASTRY, University of Surrey, United Kingdom
BAHMAN JAVADI, Western Sydney University, Australia
RAJKUMAR BUYYA, The University of Melbourne, Australia

Traditional, slow and error-prone human-driven methods to configure and manage Internet service requests are proving unsatisfactory. This is due to an increase in Internet applications with stringent quality of service (QoS) requirements. Which demands faster and fault-free service deployment with minimal or without human intervention. With this aim, intent-driven service management (IDSM) has emerged, where users express their service level agreement (SLA) requirements in a declarative manner as *intents*. With the help of closed control-loop operations, IDSM performs service configurations and deployments, autonomously to fulfill the intents. This results in a faster deployment of services and reduction in configuration errors caused by manual operations, which in turn reduces the SLA violations. This paper is an attempt to provide a systematic review of How the IDSM systems manage and fulfill the SLA requirements specified as intents. As an outcome, the review identifies four intent management activities which are performed in a closed-loop manner. For each activity, a taxonomy is proposed and used to compare the existing techniques for SLA management in IDSM systems. A critical analysis of all the considered research articles in the review and future research directions are presented in the conclusion.

## 1 INTRODUCTION

Emergence of 5G from nascency to a new global wireless standard is making significant improvements in the current Internet services, such as mobile broadband. It is also empowering the development, deployment and delivery of new services, for example, smart factories, logistics, remote surgery, precision agriculture and many other applications with low latency requirements. By supporting wide range of applications across various verticals, such as academia, medicine, industry and agriculture; 5G will be driving the global growth and has been predicted to have $13.1 Trillion of global economic output by 2035 [16]. To capitalize on such demand, communication service providers (CSPs) must offer services that can cope with associated increase in data generation and consumption. This compels them to expand and modernize their methods to deploy and operate networks and services. This includes the adoption of multi-domain, elastic and scalable solutions characterizing clouds, such as network function virtualization (NFV) [87] and software defined networks (SDN) [58]. SDN and NFV brought many benefits to simplify network services and management, but all innovation took place at the deployment level. Consequently, service

*This is the corresponding author

Authors' addresses: Yogesh Sharma, yogesh.sharma@uregina.ca, University of Regina, Saskatchewan, Canada; Deval Bhamare, d.bhamare@samsung.com, Samsung Research (SRUK), United Kingdom; Nishanth Sastry, n.sastry@surrey.ac.uk, University of Surrey, United Kingdom; Bahman Javadi, b.javadi@westernsydney.edu.au, Western Sydney University, Australia; Rajkumar Buyya, rbuyya@unimelb.edu.au, The University of Melbourne, Australia.

design and implementation are still human-driven, with system/network architects or engineers interpreting service requirements and implementing them. This is termed as a *"person+process"* approach, which is imperative or prescriptive in nature where the system is required to be told *'how'* to realize the service request [115]. However, the increasing demand of applications with stringent quality of service (QoS) requirements (high availability, throughput, security and low latency) calls for human-free service deployment to achieve desired results. It is, therefore, imperative that human intervention need be replaced with an autonomous approach to manage the service life-cycle.

Driven by such requirements and challenges, Intent-driven service management (IDSM) has been proposed with a goal of transition from traditional policy-based *"person+process"* operations model to zero-touch autonomous model [115]. With intent-driven interactions, users/service-providers express their service expectations and business objectives in a declarative manner without expressing *'how'* they should be achieved. Hence, an intent is defined as a *declarative expression* describing *what a user desires to achieve* instead of *how it should be achieved.* Once an intent is specified, closed control-loop operations of the IDSM system will work in an autonomous manner to meet the service level agreement (SLA)[1] requirements of a service request. However, the enablement of IDSM systems need complex and multi-layered arrangement including intent handlers (IH) and service orchestrators and controllers managing the resources of multiple domains/sub-systems ranging from the edge, CSP and cloud (Section 2.2). All these components need to interact, coordinate and work together in a closed loop manner towards the fulfillment of intents. Since IDSM systems are in their infancy, there is limited knowledge about their operations and activities, raising concerns about their reliability and performance variability, which could compromise SLAs. Therefore, it is imperative to have a deep understanding of the activities an IDSM system performs in order to meet the SLA requirements and to fulfill the intents.

This study is an attempt to provide a systematic landscape of SLA-based research in IDSM systems to understand the state of the art and open challenges. It provides an insight for devising solutions that address the fundamental problems in SLA management in IDSM systems. The main contributions of the paper are as follows:

- Categorization of activities the IDSM system performs to fulfill the intents.
- A comprehensive taxonomy for SLA management in IDSM systems.
- A broad review to explore various existing methods and techniques for SLA management in IDSM systems.
- Comparison and categorization of the existing techniques.
- Identification of research gaps and open challenges in the domain of SLA management in IDSM systems based on the key observations derived from the taxonomy and survey results.

The rest of the article is organized as follows. Section 2 presents the background covering the evolution, architecture and activities of IDSM systems. Section 3 describes the motivation behind the review and provides the comparison with existing reviews on IDSM. In Section 4, we discuss the research methodology followed to conduct the review and quantitative outcomes of the methodology. Section 5 presents the results of the review covering taxonomies and analysis of the research articles. Section 6, provides the critical analysis, key observations and future research directions in the area of interest. Finally, conclusions are drawn in Section 7. Table 1 shows all the abbreviations used in this survey.

## 2 BACKGROUND

With service requirements or service level agreements (SLAs) specified as intents, intent-driven service management (IDSM) systems meet these requirements autonomously. This accomplishes by taking decisions about service design, configuration, optimization, and remediation with little or no human involvement. Because of such self-driving and self-organizing properties, IDSM systems have garnered the attention of academic and industrial researchers in the fields of networking [57] and cloud computing [99]. To facilitate the research and development (R&D) efforts in the topic of interest, this section provides the information about the background of IDSM systems covering their evolution, architecture and main activities performed for intents management.

[1]SLA is an agreement between service provider and consumers regarding QoS expectations and associated reward, if met.

Table 1. List of abbreviations used in the study

| **Abbreviation** | **Full-form** | **Abbreviation** | **Full-form** |
|---|---|---|---|
| *IDSM* | Intent-driven service management | *CSP* | Communication service provider |
| *NSP* | Network service provider | *NFV* | Network function virtualization |
| *SDN* | Software defined network | *QoS* | Quality of service |
| *AI* | Artificial intelligence | *O&M* | Operation and management |
| *ECP* | Edge cloud provider | *HCP* | Hyper-scale cloud provider |
| *IoT* | Internet of things | *VR* | Virtual reality |
| *TCO* | Total cost of ownership | *CPEX* | Capital expenditure |
| *OPEX* | Operating expenditure | *SHV* | Standard high volume |
| $I-NBI$ | Intent-northbound interface | *IBNS* | Intent based networking systems |
| *RMSO* | Resource managers & service orchestrators | *IDN* | Intent-driven network |
| *KPI* | Key performance indicator | *ACL* | Access control list |
| *PNF* | Physical network function | *ML* | Machine learning |
| *VM* | Virtual machine | *QoE* | Quality of experience |

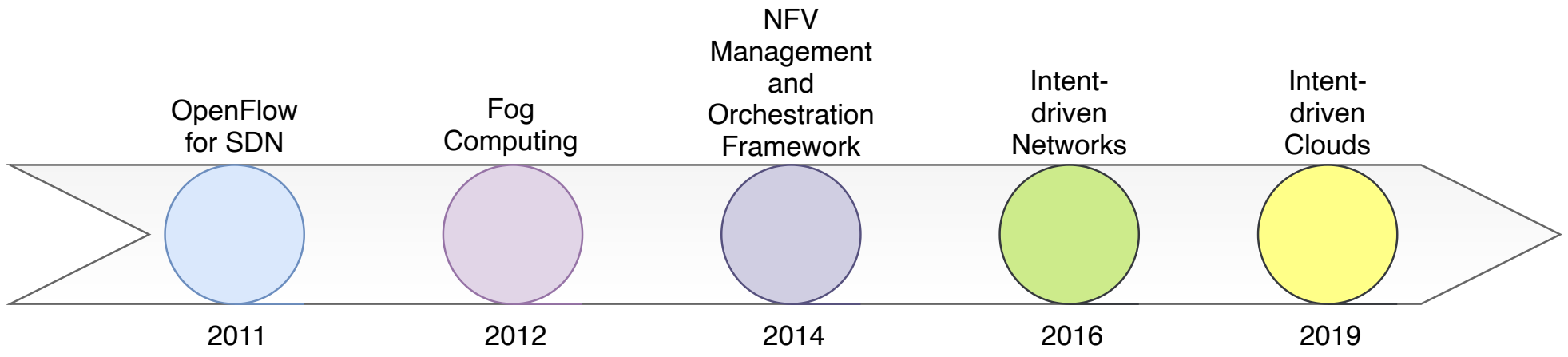

Fig. 1. Evolution of Intent-Driven Service Management systems representing the technologies and architectures that led the way to intent-driven networks followed by recent intent-driven clouds.

## 2.1 Evolution of Intent-Driven Service Management Systems

Figure 1 shows the evolution summary of IDSM systems. The steady increase in the adoption of cloud computing [27], has increased the operational and administrative complexity of computing and networking infrastructure hosting cloud services. For computing infrastructure, the complexity is dealt with significant advancements done in the field of virtualization. However, the advancement of network infrastructure (routers and switches) connecting thousands of servers hosting cloud services lags far behind. This motivated the researchers and engineers to innovate towards the softwarization of networks. With Stanford's Ethane project, efforts began in 2007 to decouple the data plane and control plane [17]. Using a centralized controller, Ethane enabled the configuration of switches and defined routing flows which led to Software Defined Networks (SDN) [58]. In 2011, OpenFlow was developed, which is a widely accepted protocol for SDNs, thereby simplifying computer networks even further [69]. SDNs and the evolution of cloud computing systems into multi-cloud/inter-cloud environments with mature interoperability enabled efforts to bring computing power closer to end users [116]. This also supported new breed of applications with low latency, real-time processing and high mobility requirements. In 2012, Cisco introduced the fog computing paradigm [13]. Fog computing components act as an intermediate layer providing compute, storage and networking services between the end user and cloud computing infrastructure. Such hierarchical arrangement aids the real-time interaction, mobility support, interoperability and scalability between end user applications and back-end cloud infrastructure. These paradigms (other is edge computing [102]) are therefore appropriate for applications that require data intensive operations as well as different processing requirements, such as Internet of Things (IoT) [46].

Networks are required to expand frequently by adding multi-specialized proprietary networking equipment to support high data volume and perform data-intensive operations. Consequently, total cost of ownership (TCO) increases in terms of capital and operating expenditures (CPEX and OPEX). In 2013, European Telecommunications Standards Institute (ETSI) started experimenting with the concept of virtualizing networking equipment as a way of taking softwarization of networking to a whole new level and reducing or eliminating the need for expensive devices [38]. In response, the concept of Virtualized Network Functions (VNFs) was introduced with networking software (control plane and data plane) hosted in VMs or containers running on Standard High Volume (SHV) servers. ETSI released its NFV Management and Orchestration framework in 2014 to provide guidelines for the deployment of VNFs to improve interoperability [87]. In the end, a decade of innovation, advances in virtualization and softwarization of computing and networking components and hierarchically deployed multi-domain paradigms became a lucrative arrangement for telecommunication industry to host their time-sensitive services. However, a more dynamic, intelligent and autonomous methods were required to configure the networks and react to the associated issues without human intervention. For this reason, in 2016, the Open Networking Foundation defined an Intent-Northbound Interface (I-NBI) and initiated the emergence of intent-based networking systems enabling the autonomous deployment and management of telco-grade applications [49]. Following the networks, in 2019, the concept of intents is adopted in the field of cloud computing system when Ericsson published an article on intent-aware cloud computing systems [99].

## 2.2 Intent-driven Service Management System Architecture

Figure 2 represents an abstract assembly of an intent-driven service management (IDSM) system. IH stands for intent handler and RMSO stands for resource manager and service orchestrator. IH is an important component of IDSM system. It is defined as *"a function which receives the intent, takes decision if and how to act, dispatches operational actions and report progress back to the source of the intent."* The IDSM system is built by assembling the IHs in a tree-like hierarchical structure sharing parent-child relationship with each other. IHs at different levels are divided into operational layers to represent the diversity of user types and roles. There can be $n$ number of operational layers and each layer can have one or more IHs. RMSO represents the domain/sub-system responsible for providing virtual and physical resources to fulfill the intents. An IH can have IHs and/or RMSO as children.

Based on the arrangement shown in Figure 2, a reference architecture of multi-layered IDSM system is shown in Figure 3. The architecture consists of 3 operational layers i.e., business, service and infrastructure. Infrastructure layer consists of three self-governing domains of edge, communication service provider (CSP) and cloud. Each layer and domain has an IH [115].

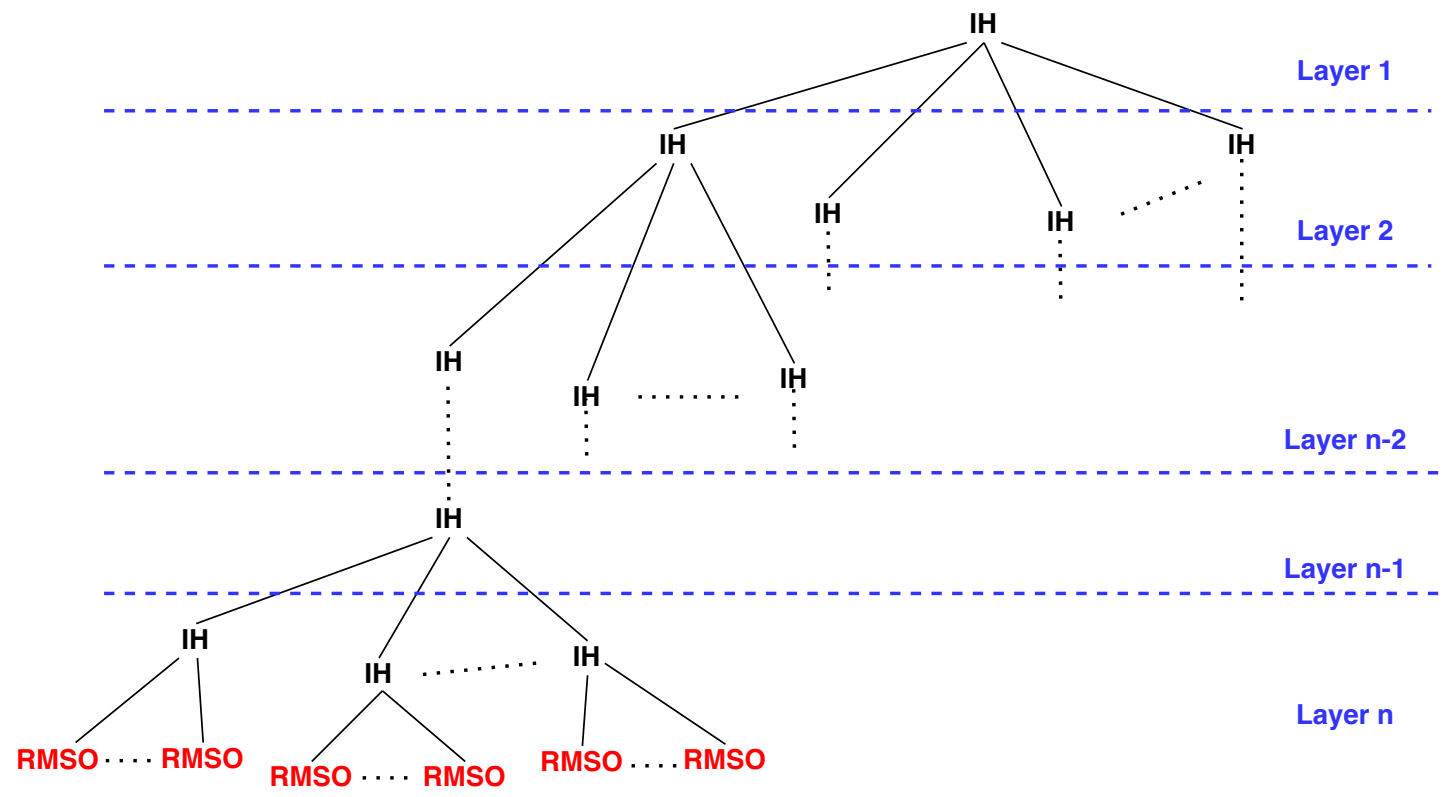

Fig. 2. Layered assembly of Intent-driven Service Management System representing hierarchical arrangement of intent handlers (IH) and resource managers and service orchestrators (RMSO).

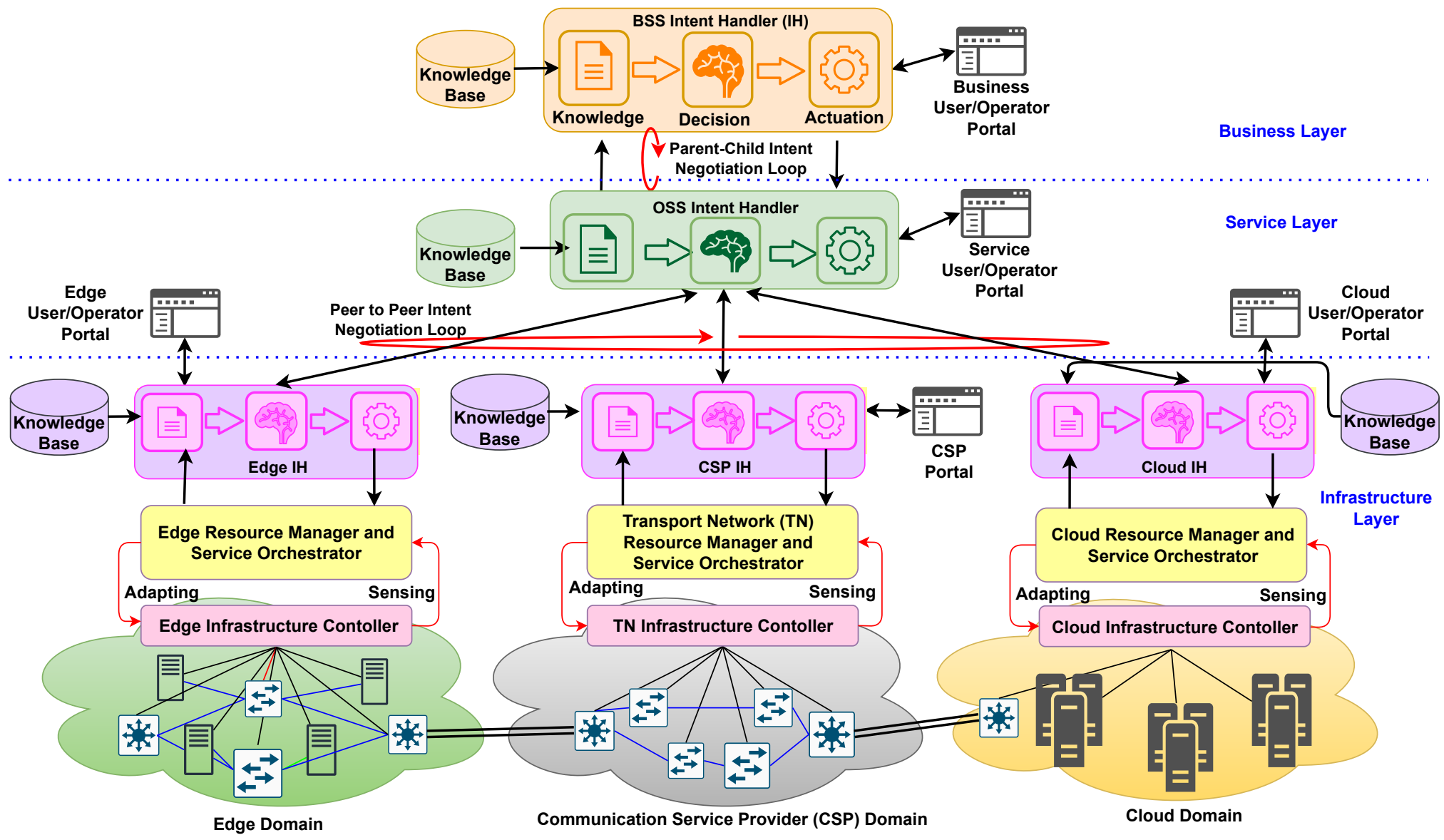

Fig. 3. Multi-layered IDSM system architecture consisting layered arrangement of intent handlers, control loops and autonomous domains of edge, CSP and cloud.

(1) Business layer IH handles the business-intents representing the functional requirements of a business user, for example, delivery of an application with customized features as defined in SLA.
(2) Service layer IH handles the intents representing the objectives of service user or provider to support business intents. The service layer intents can have more specific non-functional requirements, such as latency, bandwidth and availability.
(3) Infrastructure layer IH (domain specific IHs) handles the intents of resource users or providers. They interact with RMSO to provision and allocate resources to the service request specified as intents.

IH drives the knowledge about the intent processing operations (Section 5.1.4) from the associated knowledge-base and/or other IHs and users. Knowledge-base stores the data representing human's experience and judgment skills. ML and AI enabled IHs use the information from the knowledge-base to drive their intelligence to perform complex decision-making. It is required to design, deploy and maintain the service management operations to fulfill the intents. IHs of different layers interact with each other and with RMSOs of various domains/sub-systems by using intent-driven interfaces i.e., intent APIs in a closed loop manner. Alike operational layers, there can be more or less than 3 domains and each domain can have multiple sub-systems owned by single or multiple service providers. Each sub-system will have an associated RMSO and infrastructure controller. Intents can be originated either directly from the user input through portals or from other IHs in the hierarchy.

Upon receiving an intent, IH performs a preliminary assessment by checking its ability to fulfill the intent by using its knowledge base. If not, intent gets rejected and intent-negotiation (Section 5.1.4) starts by proposing alternative intents to the intent specification entity [100]. If yes, IH defines the goals for its child IHs by decomposing the received intent into sub-intents. With each decomposition, an intent gets enriched with the service design and configuration parameters required for the service deployment. The cycle of intent-decomposition keeps repeating in a top-down manner until the decomposed intents reach IHs local to RMSOs of the required domains/sub-systems (IHs at the layer $n$ in Figure 2). Upon receiving the request, the respective RMSO checks the availability of the required resources by probing the corresponding infrastructure controller. If the required resources are available, resource configuration parameters are forwarded to the infrastructure controller for service deployment (Section 5.2). The fulfillment of an intent is ensured throughout its lifetime in a closed-loop manner by performing continuous monitoring (Section 5.3) and remediation (Section 5.4).

On the contrary, if enough resources are not available, RMSO shares the information about the available resources with the local IH. By using the information, IH composes the alternate intents with changed or degraded service requirements. The alternate intents are used to initiate the intent-negotiation either with the parent IH or with peer IHs. If the negotiation is successful and an alternate intent is accepted then service is deployed. Alternatively, the current IH pushes the alternate intents to its parent IH in the hierarchy. The parent IH again performs the intent-composition and negotiation with its parent and peer IHs to decide about the acceptance or rejection of alternate intents. This process of intent-composition and negotiation keeps repeating in bottom-up manner until either an alternate intent is accepted or the IH where the intent was specified at fist place is reached. This is where the final decision on intent rejection or acceptance takes place and user is notified and/or asked to re-specify the intent. Together all these inter-connected components of multiple layers provide an autonomous, optimal and reliable service delivery and management at a scale and velocity which in not achievable in traditional human-driven service management systems.

## 2.3 Activities for Intent Management

In intent-driven service management (IDSM) systems, a user specifies the intents. The system adapts and changes by itself to achieve the desired results without human intervention. The journey from defining an intent to its fulfillment involves four activities that IDSM systems perform to satisfy the intent owner's service requirements (Figure 4) [125]. In this section, we are defining these activities in brief. However, all these activities are explored in depth in Section 5.

(1) *Intent Specification and Translation:* The IDSM system accepts service requirements from users specified with high-level of abstraction as 'intents'. It converts them into system design and configuration instructions with the help of an intent handler (IH).

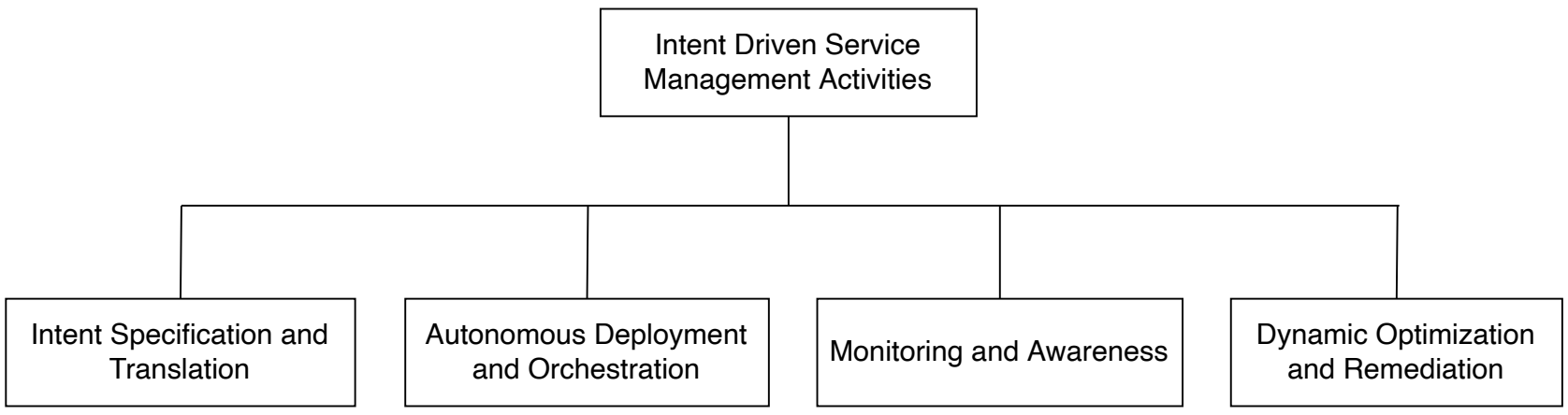

Fig. 4. Activities the intent-driven service management system performs to fulfill the intents.

(2) *Autonomous Deployment and Orchestration:* Resource managers and service orchestrators (RMSO) accept the service design and configuration instructions generated by the IHs. The required changes are performed autonomically across the software/hardware resources of multiple domains/ sub-systems to fulfill the intents.

(3) *Monitoring and Awareness:* The goal of this activity is to measure the satisfaction level of the intents. During this activity, the telemetry data is collected to evaluate the current state of the system and correlate it with the desired state of the system. It is to identify any performance deviation or anomaly that can impact the fulfillment of an intent.

(4) *Dynamic Optimization and Remediation:* If a performance deviation is identified during monitoring and awareness activity, the IDSM system takes the corrective actions by performing internal service and resource optimizations and re-configurations. It is to safeguard the fulfillment of intents or by notifying the end-users about its inability to fulfill the intents.

Ideally, IDSM systems perform all the four activities. However, during this survey, specific solutions are seen addressing fewer activities and still be the part of an IDSM solution (Section 5). In the next section we discuss the motivation behind this systematic review.

## 3 MOTIVATION BEHIND THE REVIEW

It has been observed that there are very few detailed surveys of intent-driven service management (IDSM) systems available in the literature. Table 2, summarizes the existing important survey works on the related topic and compares them with our survey.

Table 2. Comparison of Available Surveys in Intent-driven Service Management with this Survey

| Authors | Systematic Review | Evolution & Origin of IDSM | Activity Distribution of IDSM | Taxonomy for IDSM | Comparative Analysis of IDSM Solutions | Key Observations & Challenges |
|---|---|---|---|---|---|---|
| Zeydan et al.[133] | | ★ | | | | ★ |
| Pang et al.[86] | | ✓ | | | | ★ |
| Wei et al.[125] | | | ✓ | | | ★ |
| Mehmood et al.[72] | ✓ | ✓ | | | ✓ | ★ |
| Leivadeas et al.[62] | ✓ | | ✓ | ★ | ✓ | ★ |
| This Survey | ✓ | ✓ | ✓ | ✓ | ✓ | ✓ |

*Note:* ✓denotes the broad discussion on the respective issue.
*Note:* ★ denotes the partial discussion on the respective issue.

The existing surveys are not systematic reviews and performed in an ad-hoc manner except Mehmood et al. [72] and Leivadeas et al. [62]. Hence, this survey is best placed against these two systematic reviews. All of the considered surveys in Table 2 are limited to the networking field i.e., intent-driven networks (IDN). Additionally, these surveys do not discuss the activities that must be performed during the lifetime of an intent except [62]. Furthermore, they do not or partially provide a taxonomy classifying the methods and solutions for IDSM. Moreover, none of the existing surveys present a critical analysis of the existing IDSM solutions and highlight their limitations.

Table 3. Activity wise research questions answered in this systematic review.

| Activity | Research Questions |
|---|---|
| Intent Specification and Translation | 1. What are the different types of intents?<br>2. What are different languages to express or define the intents?<br>3. What are different intent stakeholders?<br>4. What are various attributes an intent can have?<br>5. What are various steps and methods/techniques to process an intent into a system adaptable form? |
| Autonomous Deployment and Orchestration | 1. What are the service level agreement (SLA) parameters of interest to intent stakeholders?<br>2. What are various SLA-based network and resource provisioning and allocation techniques used to realize the translated intents? |
| Monitoring and Awareness | 1. What are various performance challenges or bottlenecks that can breach the constraints of intents?<br>2. What are the available methods to monitor the compliance of intents?<br>3. What are various Key Performance Indicators (KPIs) used by performance monitoring methods?<br>4. What are the available methods to predict the dynamics of performance changes across the multiple layers of intent-driven service management (IDSM) systems? |
| Dynamic Optimization and Remediation | 1. What are various intention guarantee management methods?<br>2. What are the available system optimization and refinement methods require to safeguard the fulfillment of intents against any anomaly detected or predicted during monitoring and awareness activity? |

The need of addressing these shortcomings motivated us to conduct a systematic review presented in this article. Besides constructing the taxonomies and comparing the existing IDSM solutions, we performed a critical analysis of the existing literature and made a few key observations. This results in the identification of research gaps and provides the future directions to the researchers working to improve the IDSM systems.

The following section presents the details of the research methodology used to carry out this systematic review. The research methodology is based on the guidelines for performing systematic literature reviews provided by Kitchenham et al. [56].

## 4 RESEARCH METHODOLOGY

This systematic study is performed following a multi-stage research methodology, including the selection of search keywords to retrieve information from various online venues, formation of review methodology and analysis; and management of retrieved information by using review methodology. This section gives the information about all the components of the multi-stage research methodology and its outcomes.

### 4.1 Research Questions

The main goal of this systematic review is to understand the current research and development trends focusing on SLA management in IDSM systems and to identify the open challenges and research gaps in the existing research. A list of IDSM activity wise (Figure 4) research questions drafted to drive this review is provided in Table 3.

### 4.2 Sources of Information

To identify the articles on the topic of interest, electronic database search using different search keywords ( Table 4) is performed. Various research articles and reports are retrieved from the different venues, such as conferences, journals, master and PhD thesis, magazines and white papers (technical reports and industry research work). Following is the list of searched electronic databases.

Table 4. Various search keywords, period and venue types used to retrieve research articles for the review.

| Search Keywords | Period | Venue Type |
|---|---|---|
| Intent based systems<br>Intent driven/based networks (IDN)<br>Intent driven/based clouds/cloud computing<br>Intent Specification<br>Intent Decomposition<br>NorthBound Interface (NBI)<br>Intent North Bound Interface (I-NBI)<br>Intent Deployment in Networks/Clouds<br>Intent Orchestration in Networks/Clouds<br>Intent Monitoring in Networks/Clouds<br>Intent Optimisation in Networks/Clouds | 2016-2021 | Conferences<br>Journals<br>Technical and Industrial Reports<br>White Papers<br>Master and Ph.D. Thesis |

- IEEE Xplore - https://ieeexplore.ieee.org/Xplore/home.jsp
- ACM Digital Library - https://dl.acm.org/
- ScienceDirect - https://www.sciencedirect.com/
- Wiley Online Library - https://onlinelibrary.wiley.com/
- Springer - https://link.springer.com/
- Taylor & Francis Online - https://www.tandfonline.com/
- Google Scholar - https://scholar.google.com/
- Tmforum - https://www.tmforum.org/

## 4.3 Search Criteria

Table 4 describes the search keywords used to retrieve the research articles from different e-resources as discussed above. The keyword 'intent' is included in almost all the searches and found in the abstract of every searched article. We performed a careful database search to ensure the completeness of our study. Even so we couldn't get some of the research works during the predefined search method. This is due to the non-availability of search keywords in the abstract because of the synonyms being used. We retrieved some of those missed research articles by using the references of the identified papers (snowball technique). Articles published from 2016 to 2021 are considered in this review.

## 4.4 Inclusion and Exclusion Criteria

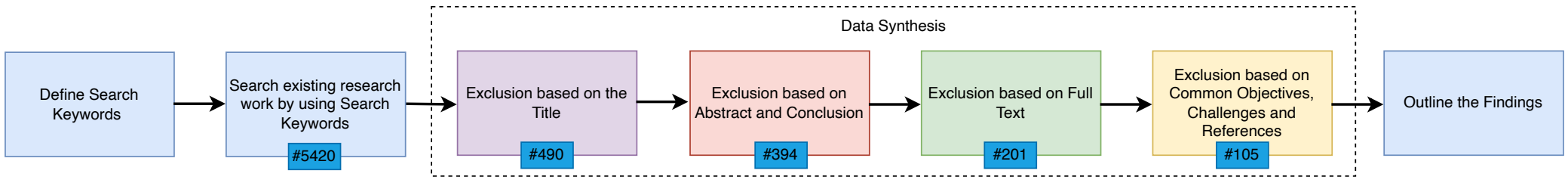

Fig. 5. Review methodology representing different stages to carry out the systematic review

Figure 5 shows the multi-stage review methodology representing inclusion and exclusion criteria used in this systematic review. By using the search keywords, we obtained 5420 research articles in total from the digital libraries. In the first stage of data synthesis, the irrelevant articles are excluded if word 'intent' is not present in the titles. As a result, 490 research articles are obtained on which the second stage exclusion process is performed by using their abstracts and conclusions. In the second stage, the articles are considered only if their focus of study is intent-driven service management (IDSM) systems. In the literature, voice command systems, such as Apple's Siri, Amazon's Alexa and

Table 5. Data extraction guidelines representing data items extracted from all research articles.

| Data Item | Description |
|---|---|
| Bibliographic Information | Author, year, title, source of the article |
| Type of the article | Journal, conference, thesis, symposium, technical report |
| Study Classification | Type of article research article or survey paper, targeted domain, publication institution |
| Study Context | What are research focus and aims of the work? |
| What are intent-driven service management systems? | It explicitly refers to activities of intent-management systems and their attributes. |
| Critical Analysis | This refers to the identification of strengths and weaknesses of each research work. |
| Study Findings | Major finds or conclusions drawn from the primary study. |

Google and autonomous cars, predicting the intents of other cars and pedestrians are also termed as intent-driven systems. Articles related to such topics are excluded and 394 articles remain. In the third stage, a thorough study of remaining articles is performed while looking for the answers to the research questions in Table 3. In this state, the number of articles is reduced to 201 based on the analysis of their full text. These articles are further filtered to 105 in the fourth exclusion stage based on their overlaps and common objectives (found in the papers from the same research group). Following the rigorous analysis of 104 articles, findings are summarized as taxonomies and tables; and presented in Section 5 and Section 6 of this paper.

### 4.5 Data Extraction

Table 5 displays the guidelines for data extraction from all the 105 research articles included in this review. Various problems were faced regarding the extraction of suitable data, for example, information is missing or not clearly available in the article. To get clarification about the missing information, we contacted the authors of the respective research articles. While extracting the data, all the authors of this review communicated and held meetings regularly and performed an in-depth analysis of the research works as described below.

- First author extracted and analyzed the data from 105 research articles.
- Other authors cross-checked the results to check the consistency of the extracted data.
- Conflicts occurred during cross-checking were resolved during the meetings.

### 4.6 Quantitative Analysis of Research Methodology

Figure 6 depicts the quantitative analysis of 105 research articles considered in this review. In Figure 6.1, it has been observed that 68% of total research articles are published during the time period of 2020-2022 with 2022 having the biggest share of 35%. This shows the increasing interest of researchers in intent-driven service management (IDSM) systems. Figures 6.2 and 6.3 represent the publication venue and institution wise distributions of the research articles. As depicted, most of the research is published in conferences (65%) followed by journals (29%). Whereas, publications coming out of academic institutions are the major contributors (53%) followed by the articles published in collaboration between academic institutions and their industrial partners (29%). Figure 6.4 is the collective representation of number of publications vs venue type and year of publication. It can be seen that the number of publications in journals are increasing consistently since 2018. This represents that the research in IDSM systems is progressing and the quality of solutions is improving and maturing, which is analyzed and explained in the following section.

## 5 A TAXONOMY

In Figure 4, we identified four activities the intent-driven service management (IDSM) systems perform to fulfill the service level agreement (SLA) requirements of the intents. In this section, a thorough study of each activity is performed and corresponding taxonomies and formal definitions are provided. This section also compares the solutions for IDSM systems from the literature.

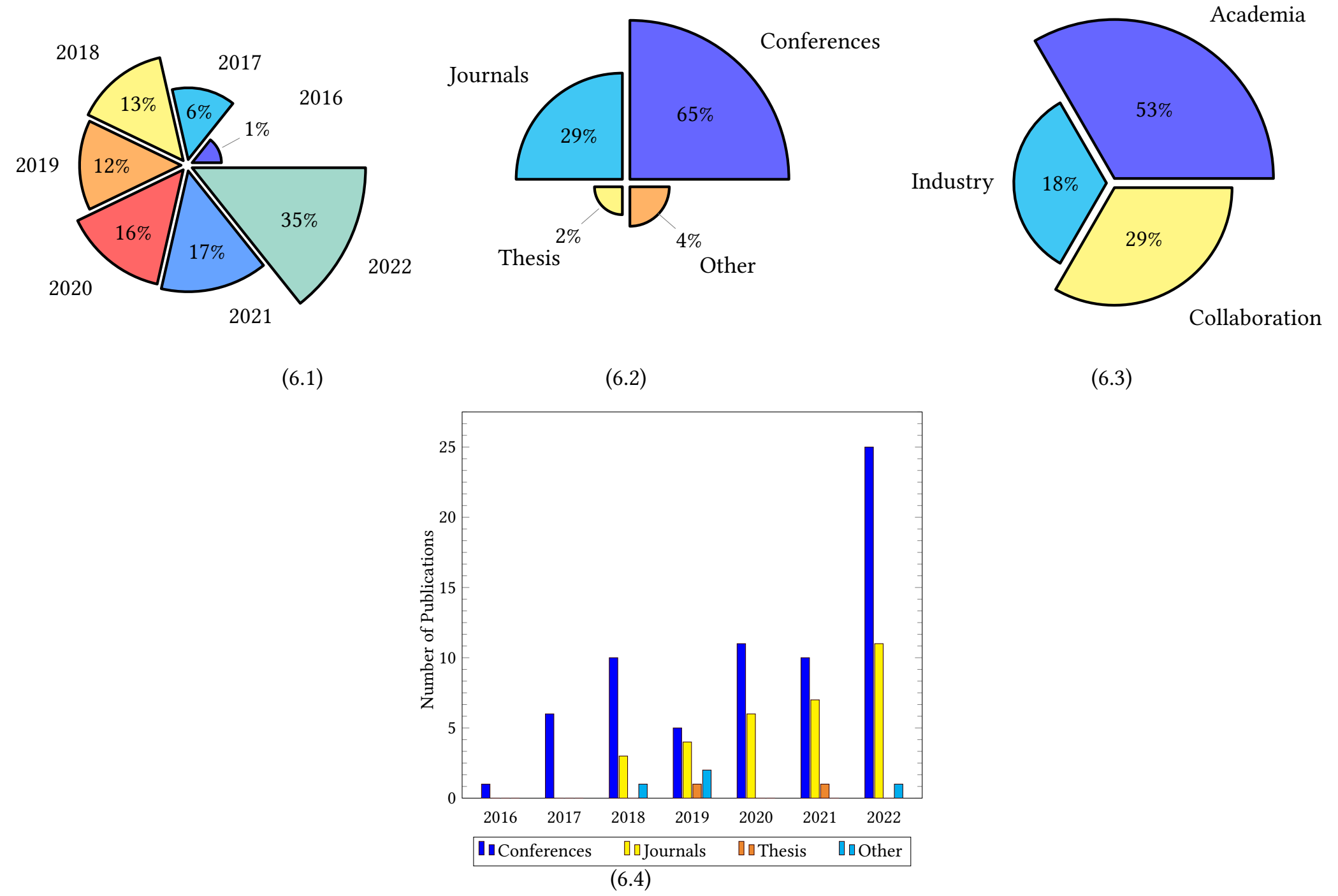

Fig. 6. Quantitative analysis of research methodology showing distribution of research articles according to (i) year of publication (ii) venue of publication (iii) publishing institution (iv) Comparison of number of publications vs venue types vs year of publication.

### 5.1 Intent Specification and Translation

In this activity, intent handler (IH) captures the high-level intents and converts them to required system design and policies. Figure 7 shows the taxonomy for intent specification and translation representing various components of an intent, such as intent types, attributes, specification languages, intent processing methods and languages of configuration output after processing an intent. Each component is discussed in the following sections along with their sub-components and suitable examples. The analysis of various methods and solutions addressing intent specification and translation is presented in Table 6.

*5.1.1 Intent Specification:* It is an act of stating/describing the intents representing expected outcomes/results in the form of high-level service requests. An intent can have multiple stakeholders i.e., service users and providers, and can be specified by using a (1) Formal or (2) Informal language.

- *Formal Languages:* Languages with precise syntax and semantics are called formal languages. Intents specified using formal languages needs less or no pre-processing before being fed to the intent handler for further processing (Section 5.1.4). Some of the formal languages frequently used to specify intents are: JSON [32, 65, 118, 121], XML [24, 35, 54], Scala [41], NEMO [117] and SPARQL [20].

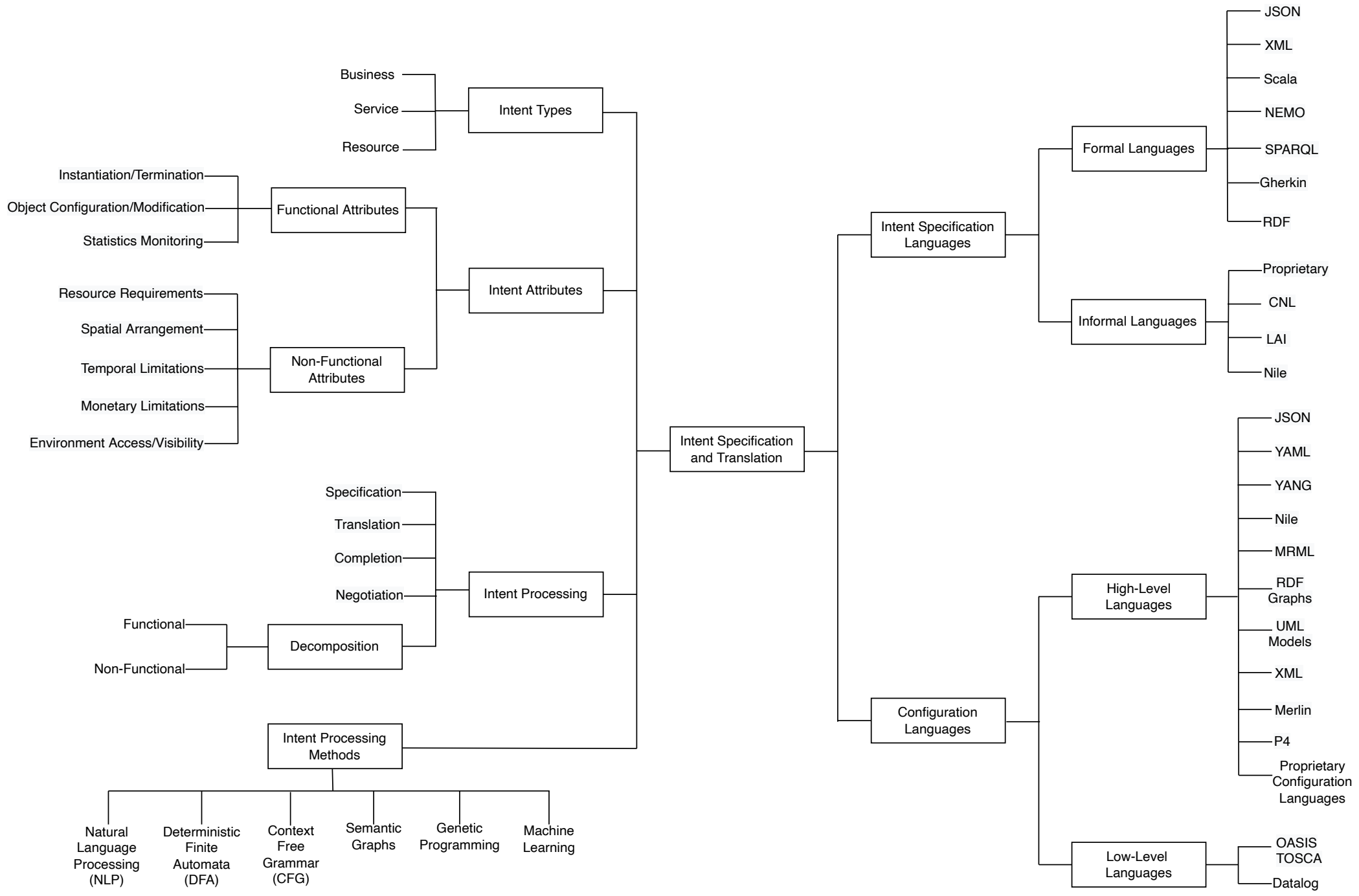

Fig. 7. Taxonomy for Intent Specification and Translation Activity

- *Informal Languages:* These languages are either Controlled Natural Languages (CNL) used by the humans in daily routine or a blend of formal and CNL also called 'pseudo code'. Informal languages are more solution/user specific languages with a loosely defined syntax that can change according to the use case. Intents specified using informal languages are tend to have ambiguities. An intermediate processing system is required to resolve such ambiguities before they can be used as input to an intent handler. Apart from CNL [6, 55, 97, 127] and proprietary languages [50, 59, 110], other informal languages commonly used to specify intents are: Language for Access Control List Intents (LAI) [114] and Nile [89, 92, 123].

*5.1.2 Intent Attributes:* Intent Attributes provide the key information about the characteristics of an envisioned service request specified as an intent. They are of two types: (1) Functional and (2) Non-Functional attributes.

- *Functional Attributes:* Functional attributes represent what a service or system is expected to do or perform to fulfill the objectives of an intent. In Figure 8, keywords 'Features' and 'Topology' represent the functional attributes illustrating the need to connect site X and Y by deploying a link between them. Based on the characteristics identified, an intent can have three classes of functional attributes: (1) Instantiation/termination, (2) Object configuration/modification and (3) Statistics monitoring.
  Instantiation/termination attributes represent the need to start or stop a service instance. Object configuration/modification attributes express the requirement of changing the configuration of an instance of a service. Statistics monitoring attributes represent the demand for the collection of telemetry data. Tsuzaki et al. [117], Riftadi et al. [92], Esposito et al. [37], Chung et al. [24] and Meijer et al. [75] are some of the works with intents covering both

Table 6. Summary of existing works considering Intent Specification and Translation Activity Taxonomy

| Reference | Intent Specification & Translation | | | | | | Reference | Intent Specification & Translation | | | | | |
|---|---|---|---|---|---|---|---|---|---|---|---|---|---|
| | Intent Type | Intent Attributes | Specification Language | Intent Processing | Intent Processing Method | Configuration Language | | Intent Type | Intent Attributes | Specification Language | Intent Processing | Intent Processing Method | Configuration Language |
| [110] | S, R | OC, I/T, RR | Pro | C, D | NG | Pro. | [97] | B, S, R | I/T, SA, EA | CNL | D | SG, ML | Pro. |
| [117] | B, R | OC, I/T, RR | NEMO | D | NG | Pro. | [2] | B, S, R | OC, SA, RR, TL, EA | Pro. | D | SG | Pro. |
| [50] | S | OC, EA | Pro. | D | SG | Datalog | [6] | B, S | OC, I/T, SA, EA | CNL | D | NLP | NG |
| [106] | B | OC, I/T, SA, RR, TL, EA | JSON | NG | DFA, CFG | JSON | [65] | S | OC, SA, RR | JSON | D | ML | NG |
| [26] | B, S | OC, I/T, SA, RR | Pro. | N, D | NG | Pro. | [35] | B | OC, I/T, RR | XML | D | NG | XML |
| [32] | R | OC, I/T, SA, RR, EA | JSON | C, D | SG | JSON | [121] | B | I/T, RR | JSON | D | NG | NG |
| [118] | B, S, R | OC, I/T, SA, RR | JSON | C, D | NG | Pro. | [37] | B, S | OC, I/T, SA, EA | Gherkin | NG | NG | JSON |
| [19] | S | RR, EA | CNL | D | NLP, ML | YAML | [78] | B, S, R | I/T, RR, TL | JSON | C, N | SG | MRML |
| [132] | B, R | OC, RR, EA | XML | D | DFA, CFG | XML | [55] | B | OC, I/T, SA, RR, TL | CNL | T, C N, D | NLP | RDF |
| [30] | S, R | OC, SM, SA, RR, TL | JSON | D | NG | JSON | [122] | S, R | SA, EA | JSON | D | NG | YANG |
| [112] | S | I/T, RR, EA | JSON | NG | NG | YANG | [92] | S, R | OC, RR | Pro. | C | ML | P4 |
| [127] | R | I/T, RR | CNL | D | NLP | YAML | [3] | B, S | OC, I/T, RR, TL | JSON | N, D | SG | JSON |
| [91] | S, R | OC, I/T, RR | Nile | D | SG | P4 | [15] | S, R | I/T, SA, EA | JSON | D | NG | YAML |
| [114] | S, R | OC, SM, SA, EA | LAI | N, D | CFG | Pro | [59] | S | OC, I/T, TL, EA | Pro. | D | CFG | JSON |
| [20] | B, S | OC, I/T, SA, RR | SPARQL | D | SG | RDF | [24] | B, S | OC, I/T, SA, RR, TL | CNL | NG | NG | JSON |
| [47] | S | OC, SA, EA | CNL | NG | NLP, ML | Nile | [98] | B | OC, I/T, RR, TL, EA | CNL | C, D | DFA, CFG | JSON |
| [52] | B, S | OC, RR | Pro. | NG | NG | JSON, YAML, TOSCA | [23] | B, R | OC, RR, TL | JSON XML | NG | DFA, CFG | XML |
| [119] | S | OC, I/T, SA | Pro. | D | SG | Pro. | [4] | R | OC, RR | CNL | NG | NLP | JSON |
| [66] | B, S, R | I/T, SA, RR, TL | CNL | N | NLP, ML | RDF | [80] | S | OC, SA, EA | NG | D | NG | NG |
| [41] | B | OC, I/T, RR | Scala | D | SG | Pro. | [89] | B, S | SA, EA | Nile | NG | NG | JSON |
| [81] | B, S | OC, I/T, SA, RR | NG | D | NG | UML Models | [54] | B, S | OC, I/T, EA | XML | D | DFA, CFG | XML |
| [124] | B | OC, SA, TL, EA | Nile | NG | NG | Pro. | [67] | B, S | RR | JSON | N | SG | Pro. |
| [88] | S, R | SA, RR | JSON | NG | NG | JSON | [131] | B, S, R | OC, I/T, SA, RR | CNL | N, D | NLP, ML | JSON |
| [134] | S | SA, EA | Pro. | N, D | NG | Pro. | [43] | S | SA | Pro. | D | SG | Pro. |
| [73] | B, S | OC, I/T | CNL | NG | NG | YANG | [53] | B, S | OC, I/T, SA, RR | Pro. | NG | NG | Pro. |
| [76] | B, S | I/T, SA, EA | CNL | D | NLP | Pro. | [11] | B | OC, SM, TL, EA | CNL | N | CFG | JSON |
| [12] | S, R | OC, I/T, SA, TL | CNL | NG | NLP | JSON | [85] | B, S | I/T, RR, TL | CNL | NG | NLP, DFA | JSON |
| [33] | B, S | OC, SM, SA, RR | CNL | D | NLP | JSON | [1] | B, S | OC, I/T, RR, TL | Pro. | NG | NG | JSON TOSCA |
| [34] | B, S | OC, RR | CNL | C | NG | Pro. | [61] | S, R | OC, RR, MNL | Pro. | C | SG | TOSCA |
| [31] | B, S, R | I/T, SA, TL | CNL | NG | NG | RDF | [48] | S | OC, SA, RR, EA | CNL | C | NLP, ML | Nile Merlin |
| [42] | S, R | OC, I/T, RR | NG | D | NG | RDF | [29] | S | OC, SA, RR, EA | CNL | C | NLP, ML | JSON |
| [70] | S, R | OC | CNL | C | NLP | JSON | [44] | S, R | OC, I/T, SA, RR, EA | JSON | D | NG | JSON YAML |
| [40] | R | OC, SM, I/T, SA, RR | CNL | NG | NG | NG | [60] | S, R | OC, SA, RR | CNL | D | ML | Pro. |
| [120] | S | OC, RR, TL | JSON | N, D | NG | RDF | [9] | B, S | OC, SM | CNL | N | NG | Pro. |
| [22] | B | OC, I/T, RR | CNL | D | NG | Pro. | [14] | B, S | OC, SA, TL, RR | CNL | D | NG | YAML |
| [8] | B | OC, RR | NG | NG | NG | RDF | [94] | S | OC, I/T, SA, RR, EA | CNL | C, D | NLP, ML | Pro. |
| [10] | S, R | OC, SA, RR | JSON | D | NG | YAML | [136] | B | RR, TL | Pro. | N, D | SG | NG |
| [130] | B, S, R | SM, I/T, RR | RDF | D | NG | RDF | [63] | S, R | OC, I/T, RR | NG | N, D | NG | NG |
| [77] | S | I/T, EA | CNL | D | NLP | RDF | [135] | B, S, R | OC, I/T, SA, TL | CNL | D | NLP, ML | NG |
| [18] | B, R | OC, I/T, TL | NG | C, D | CFG, ML | JSON | [129] | B, S, R | I/T, RR | CNL | C, N, D | NLP, SG | JSON |
| [74] | B, S, R | OC, I/T, RR | CNL | D | NLP | NG | [7] | S, R | OC, SA, EA | CNL | D | NLP, ML | P4 |
| [108] | B, S | OC, SM, I/T, RR | CNL | C | NLP, ML | JSON | [75] | S | OC, I/T, RR | CNL | NG | NLP, ML | Nile |
| [113] | B, S, R | NG | CNL | N, D | NLP | JSON | [107] | S, R | OC, I/T, SA, RR | Pro. | D | SG | JSON |
| [21] | S, R | OC, SM, I/T, SA, RR, EA | Pro. | N, D | DFA | Pro. | [51] | B, S, R | OC, I/T, SA, RR, EA | CNL | NG | NG | P4 |
| [64] | B | OC, I/T, RR | Pro. | D | CFG | Pro. | [90] | B, S, R | OC, I/T, SA, EA | Pro. | C, D | CFG | Nile |
| [45] | R | OC, SA | Pro. | D | GP | P4 | [138] | B, S, R | OC, I/T, SA, RR, TL | Pro. | N | NG | NG |
| [68] | S, R | OC, I/T, SA, RR | Pro. | D | NG | JSON | [84] | S, R | OC, SM, SA, RR MNL, EA | NG | D | CFG | NG |
| [126] | S | OC,I/T, RR, TL | Pro. | N | NG | NG | [100] | B, S | I/T, RR, MNL | JSON | C, N, D | NG | JSON |

**NG:** Not given, **B:** Business, **S:** Service, **R:** Resource, **I/T:** Instantiation/Termination, **OC:** Object configuration & modification, **SM:** Statistics monitoring, **RR:** Resource requirements, **SA:** Spatial arrangement
**TL:** Temporal limitations, **MNL:** Monetary limitations, **EA:** Environment access/visibility, **T:** Translation, **C:** Completion, **N:** Negotiation, **D:** Decomposition, **SG:** Semantic graphs, **GP:** Genetic Programming
**ML:** Machine learning, **Pro.:** Proprietary, **NLP:** Natural language processing, **DFA:** Deterministic finite automata, **CFG:** Context free grammar, **CNL:** Controlled natural language instantiation/termination and object configuration/modification attributes, whereas Tian et al. [114], Davoli et al. [30] and Xie et al.[130] have intents with statistics monitoring attributes.

- *Non-functional Attributes:* Non-functional attributes represent the quantitative or qualitative constraint or parameters required to be obliged while fulfilling an intent. In Figure 8, keywords 'Latency', 'Cost', 'Availability', 'Bandwidth', 'Start' and 'Stop' timestamps represent the non-functional attributes providing configuration values and corresponding constraints for a link required to be deployed between site X and site Y. Based on the characteristics of non-functional attributes, we have divided them into five categories: (1) Resource requirements, (2) Spatial arrangement, (3) Temporal limitations, (4) Monetary limitations and (5) Environment access/visibility.
  Resource requirement attributes of an intent represents the essential compute or network resources (CPU, memory, storage, bandwidth) asked by an intent owner. Temporal and Monetary limitations are the attributes for imposing time and cost related constraints on a service request, respectively (start and stop timestamps; and cost constraints in Figure 8). Spatial arrangement attributes represent the space related constraints, for example, cloud storage service within the borders of a country is requested because of the govt. regulations. Environment access attributes are related to intents for security services, such as firewall and intrusion detection systems (IDS). Abhashkumar et al. [2] and Sköldström et al. [106] have specified all of the non-functional attributes except monetary limitations which are specified in Kuwahara et al. [61] and Sharma et al. [100].

*5.1.3 Intent Types:* In IDSM system architecture shown in Figure 3, users of each layer can specify the intents with different levels of abstraction. Hence dividing them into three categories: (1) Business, (2) Service and (3) Resource intents. This categorization distinguishes the concerns and objectives of different parties involved in the intent-handling. For reader's convenience, in this sub-section, examples of different types of intents are provided in a Controlled Natural Language (CNL) and are obtained from [130]. However, any language (Formal or Informal) can be used to specify the intents (Section 5.1.1).

- *Business intent:* It represents the objectives of the business layer users interested in the delivery of customized applications defined by service level agreements (SLAs). It includes the functional attributes associated to a product or customer management of an application with revenue and quality of experience (QoE) targets as non-functional attributes. For example, *Order an entertainment service with downlink and uplink throughput equal to 30 and 10 Gbps, respectively and latency not less than 20ms* .

- *Service intent:* It represents the objectives of the service layer users responsible of designing the services, their orchestration, activation and assurance. Service intents aim to deliver the service to business users with required functional and non-functional attributes defined in the business intents. For example, *Order a cross-domain enhanced mobility broadband (eMBB) slice from Operator Y to host an entertainment application with delivery parameters as defined in business intent SLA.*
- *Resource intent:* It represents the objectives of the infrastructure layer users which handles the provisioning and allocation of resources so that the performance and quality of service (QoS) of business and service intents are met. The functional and non-functional attributes of resource intents deals with network orchestration and virtual and physical resource management. For example, *Deliver radio-access netwok (RAN), transport network (TN) and core network (CN) sub-slices meeting the QoS parameters defined in service and business intent SLA.*

*5.1.4 Intent Processing:* An intent is required to be processed by the IHs to obtain a valid expression that can be used by resource managers and service orchestrators (RMSO) to realize the service request. Processing of an intent consists of four stages: (1) Intent translation, (2) Intent completion, (3) Intent negotiation and (4) Intent decomposition. With the execution of each stage, an intent expression becomes richer and moves closer to RMSO usable form.

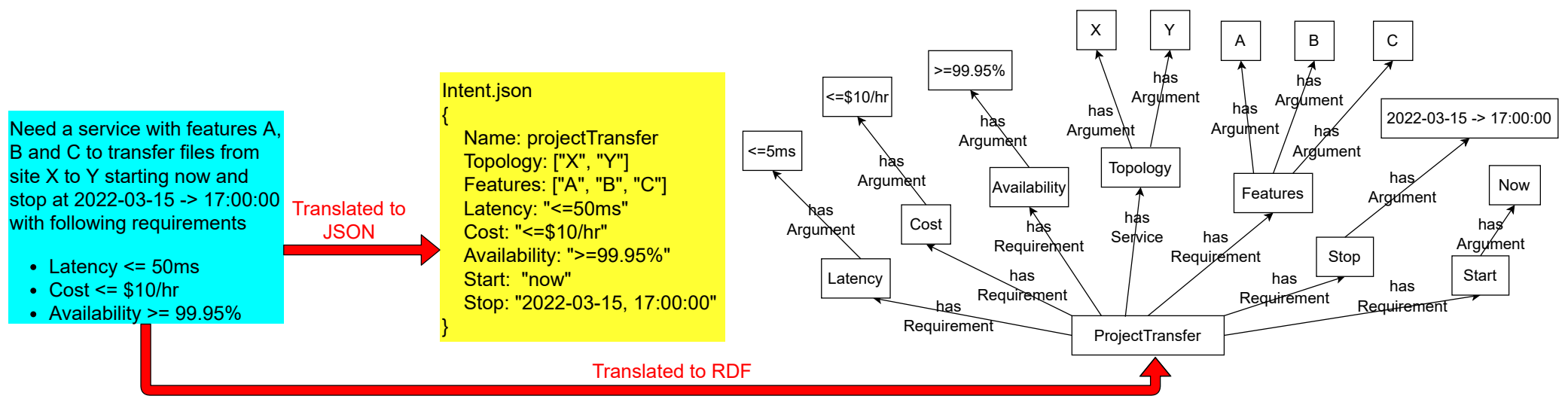

Fig. 8. Intent translation from controlled natural language (CNL) to JSON and RDF format

- *Intent Translation:* It refers to changing the notation of an intent specified by using any formal or informal language and predefined template or without template to make it interpretable by the IDSM system. Translation keeps the level of abstraction of an intent same as of specification and does not add or remove any information or details. As shown in Figure 8, an intent specified in a CNL (highlighted in blue) is translated to a template defined in JSON (highlighted in yellow) and RDF format, respectively without adding or removing any information.
- *Intent Completion:* It is a process to determine the imprecise or unknown parameters an intent expression may contain. Such parameters may be required to present in an intent format acceptable by an IDSM system. The unknown parameters can be obtained by the IHs implicitly or explicitly. While using implicit methods, one way to introduce the parameters by using the default keywords (Figure 9(a)). Which obtains their quantitative values during the process of parameter estimation amid intent decomposition [118]. The other way is to obtain such unknown parameters by integrating the service provider and user intents (Figure 9(b)) [100]. In explicit method, the IH uses a combination of iterative steps involving the intent user to ask for clarifications about the unknown parameters. This method is used by Monga et al. [78] and Kiran et al. [55] where they employed a chat-box to ask for clarification from the users about the missing parameters. Intents obtained after the completion process has the minimal information about the service design and farthest from RMSO usable form.

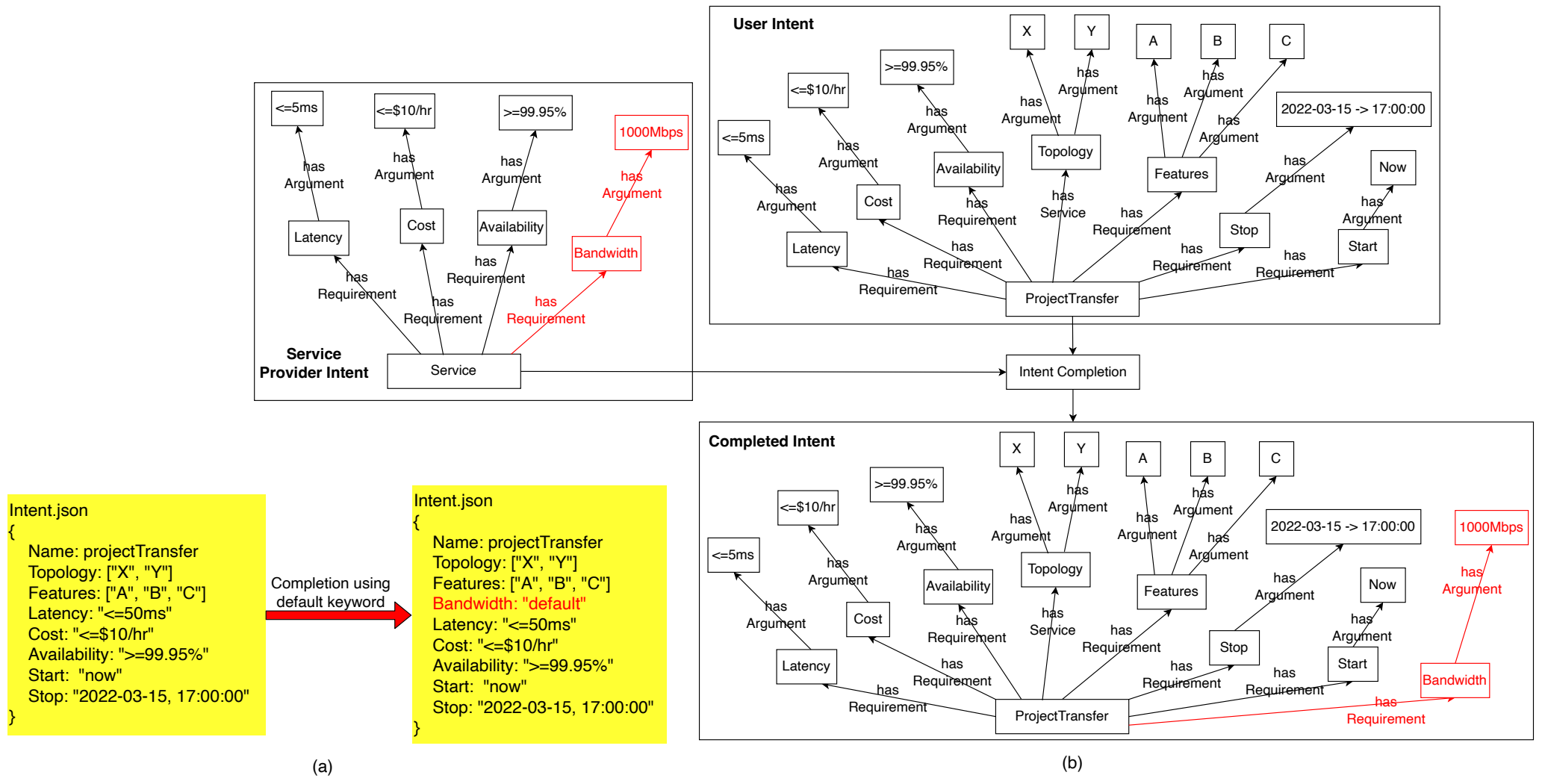

Fig. 9. (a) Intent completion is performed by adding Bandwidth parameter with "default" keyword. (b) Intent completion is performed by obtaining Bandwidth parameter by integrating service provider and user intents.

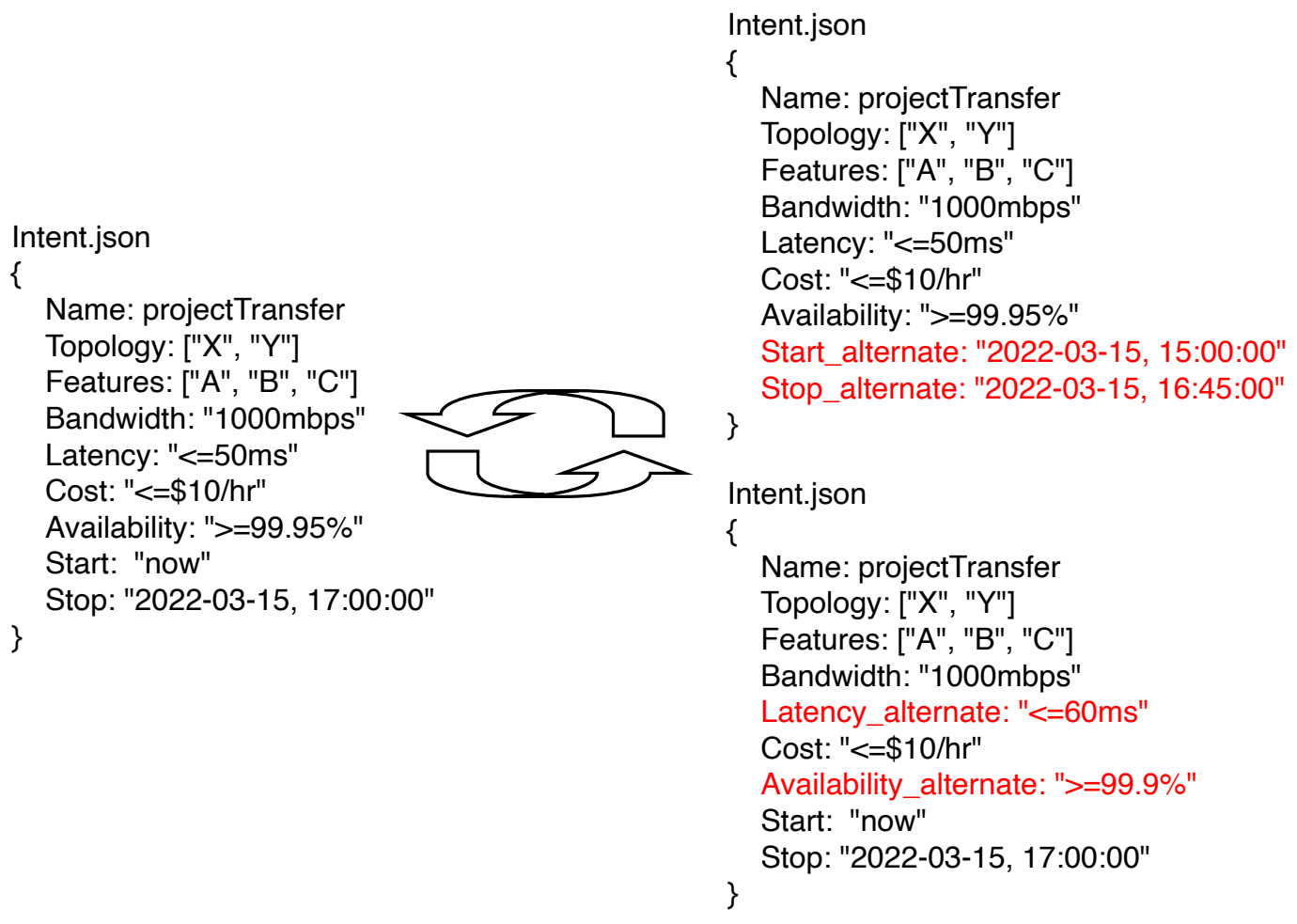

Fig. 10. Intent negotiation generating alternate solutions with relaxed temporal and resource requirements during system's inability to satisfy an intent because of resource scarcity.

- *Intent Negotiation:* It is an iterative bi-directional process of reaching an agreement between the intent user and service provider by offering alternative intents (with changed or degraded service requirements) to a given intent. This happens when the current state of the service provider cannot meet the requirements of an intent submitted by a user. Figure 10 represents the process of intent negotiation where the IH offers the two alternate solutions to the user to select from. One with changed temporal constraints (start and stop timestamps) and other with relaxed performance constraints (availability and latency). Marsico et al. [67] proposed an intent negotiation framework equipped with alternative solution selection algorithm which provides alternative solutions during resource scarcity with relaxed bandwidth, latency and availability requirements. Tian et al. [114] proposed an intent-driven access control list (ACL) updating system 'JinJing' for Alibaba's global wide area network (WAN). The system is able to detect any policy conflicts while updating the ACL configurations and provides alternate solutions to choose from to avoid such conflicts. Comer et al. [26], Teng et al. [113] and Li et al. [63] have also employed intent negotiation methods while processing an intent.

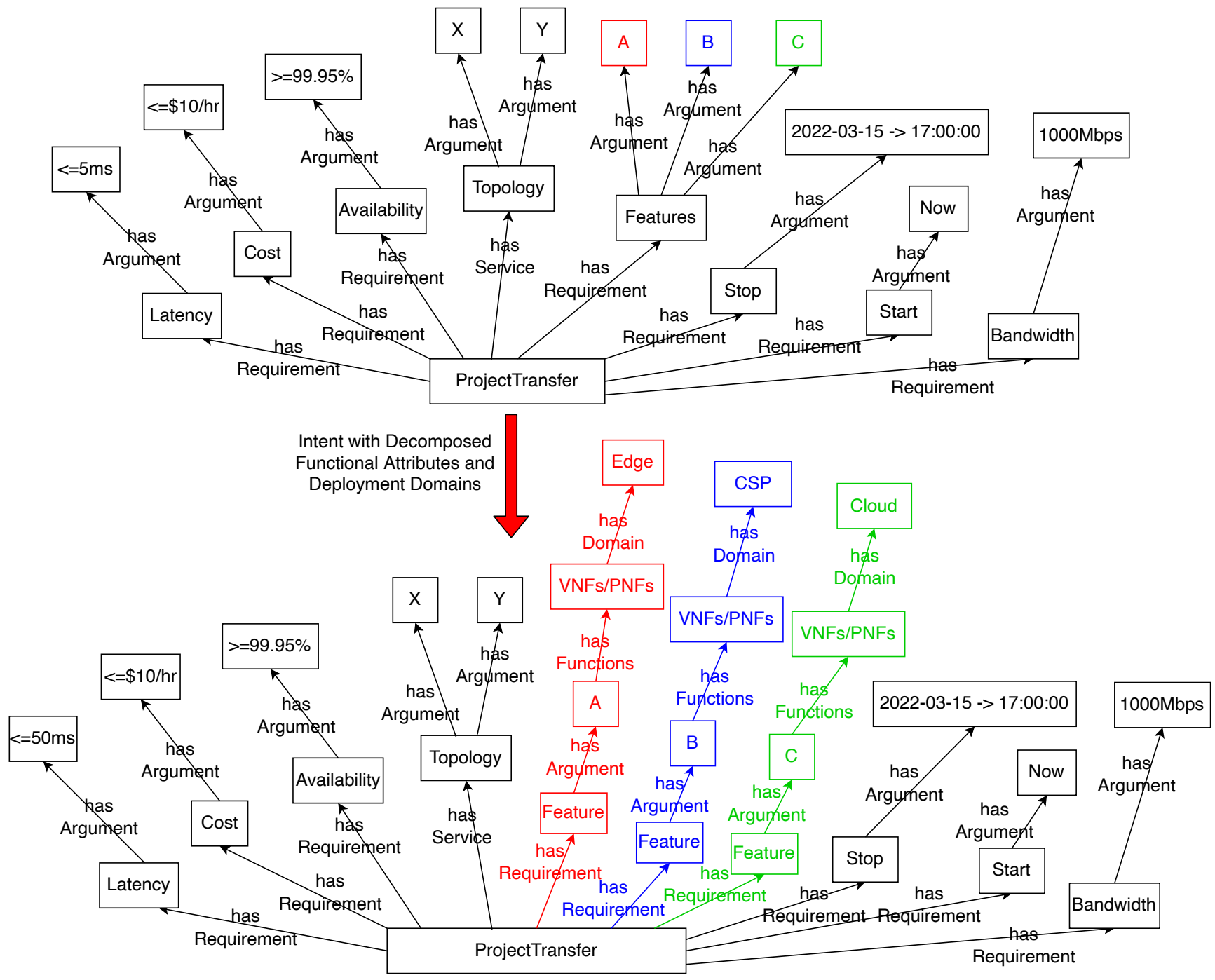

Fig. 11. Functional Decomposition of an intent identifying required VNFs/PNFs and deployment domains.

- *Intent Decomposition:* Intent decomposition breaks down a higher-level intent into sub-intents for its dissemination across different IHs or sub-systems required for its fulfillment. During intent decomposition, an intent gets enriched with the information i.e., service design and configuration parameters, required for the service deployment. Intent decomposition is of two types: (1) Functional decomposition and (2) Non-functional decomposition
  - *Functional Decomposition:* To satisfy the functional attributes of an intent, functional decomposition obtains the information about the appropriate functional components required to deploy a service. This includes the selection of virtual/physical functions, their order of deployment i.e., service chains representing interconnections between the selected virtual/physical functions. Additionally, it decides about the deployment domains/sub-systems i.e., edge, communication service provider (CSP) and cloud for the selected functional components. Figure 11 represents the functional decomposition of an intent shown in Figure 9(b) to a more precise service request. In the given figure, required virtual and physical network functions (VNFs/PNFs) are decided to host a connectivity service between X and Y with features A, B and C. Features in the present context stands for quality of service (QoS) functions similar to encryption, error detection and correction, firewall, traffic forwarding and intrusion detection system. Apart from deciding about VNFs and PNFs, domains/sub-systems where these functions will be hosted are decided. Nazarzadeoghaz et al . [81] proposed an intent decomposition framework for both functional and non-functional attributes of the intents specified specifically for provisioning and deployment of network slices. The proposed framework uses a UML based ontology (knowledge base) to get the information about required network functions and their order of deployment and corresponding configuration parameters for a network slice. Sung et al. [110], Davoli et al. [30], Chen et al. [20], Ujcich et al. [119] and Gritli et al. [43] are some of the other works addressing the challenge of functional decomposition of intents.

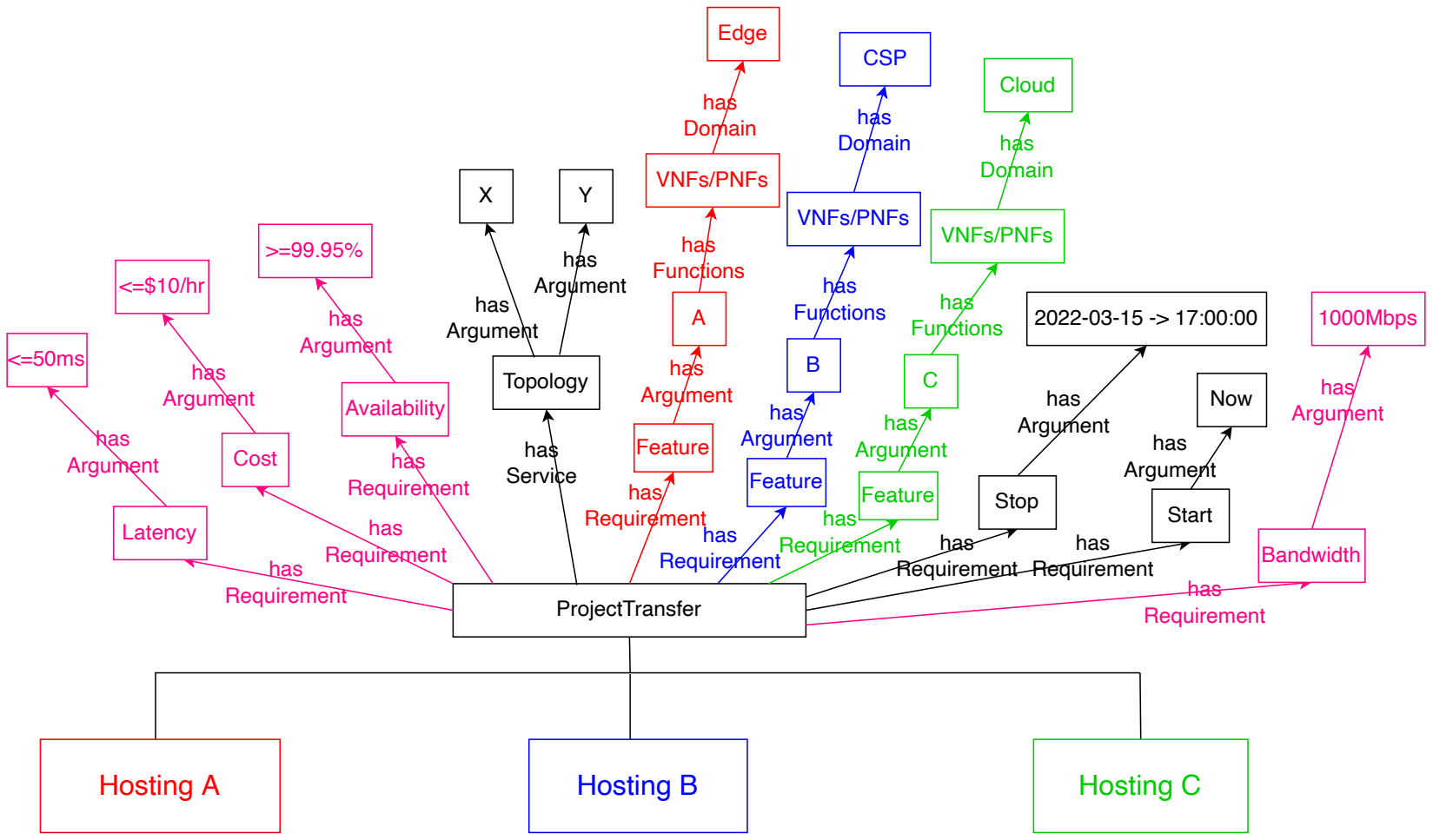

Fig. 12. Decomposition of non-functional attributes of the intent obtained after functional decomposition. It results in the break-down of the original intent into sub-intents corresponding to each deployment domain obtained during functional decomposition.

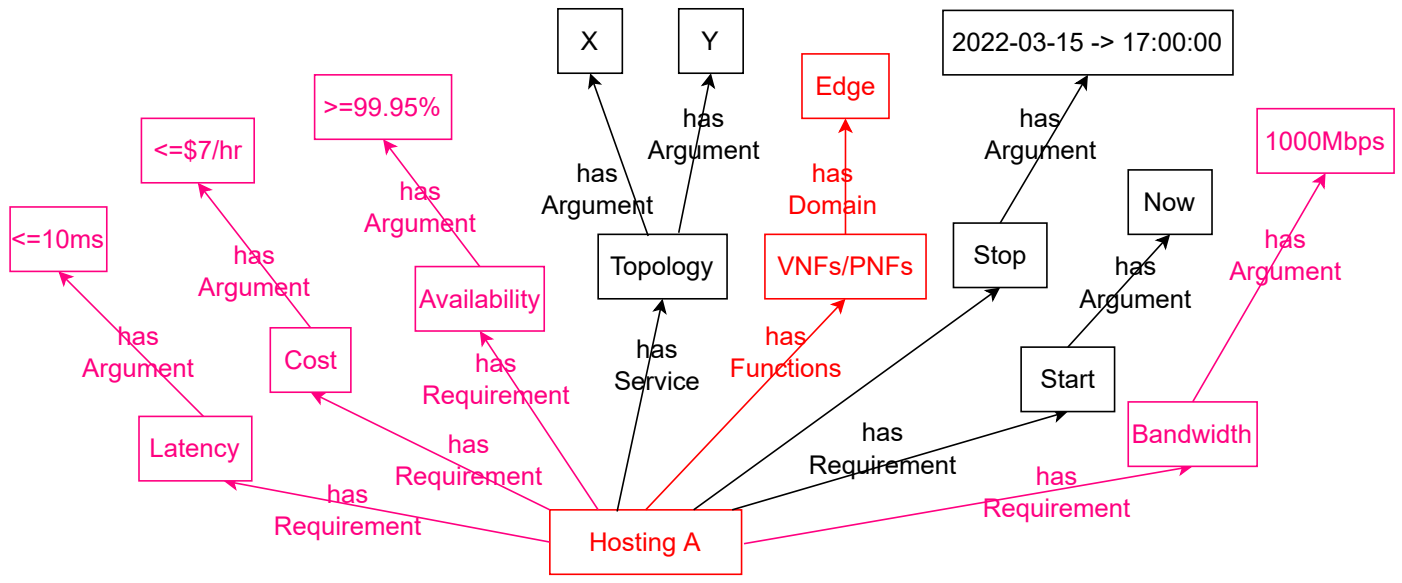

Fig. 13. Sub-intent obtained after Non-Functional Decomposition of intent to host feature A on Edge.

– *Non-Functional Decomposition:* It refers to breaking down of the performance constraints specified in an intent to sub-intents and estimation of configuration parameters for the selected physical/virtual functions during functional decomposition. Figure 12 shows the non-functional decomposition of an intent obtained after functional decomposition in Figure 11. The intent is decomposed into three sub-intents corresponding to each domain/sub-system i.e., edge (Figure 13), CSP (Figure 14) and cloud (Figure 15) hosting feature A, B and C, respectively. The non-functional attributes, such as latency, cost and availability, are decomposed according to the characteristics of the domain/sub-system hosting VNFs/PNFs where as the bandwidth remains same for all the domains/sub-systems. Lin et al. [64], Xie et al. [130] and Sharma et al. [100] are some of the works covering non-functional decomposition of intents.

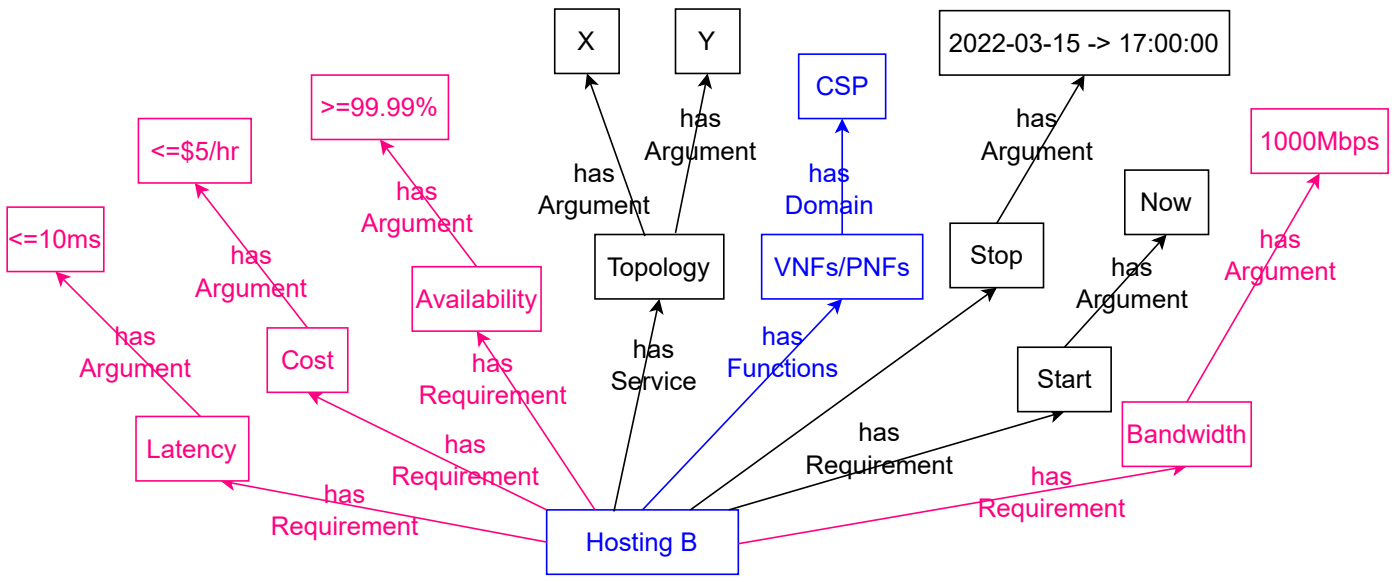

Fig. 14. Sub-intent obtained after Non-Functional Decomposition of intent to host feature B on CSP.

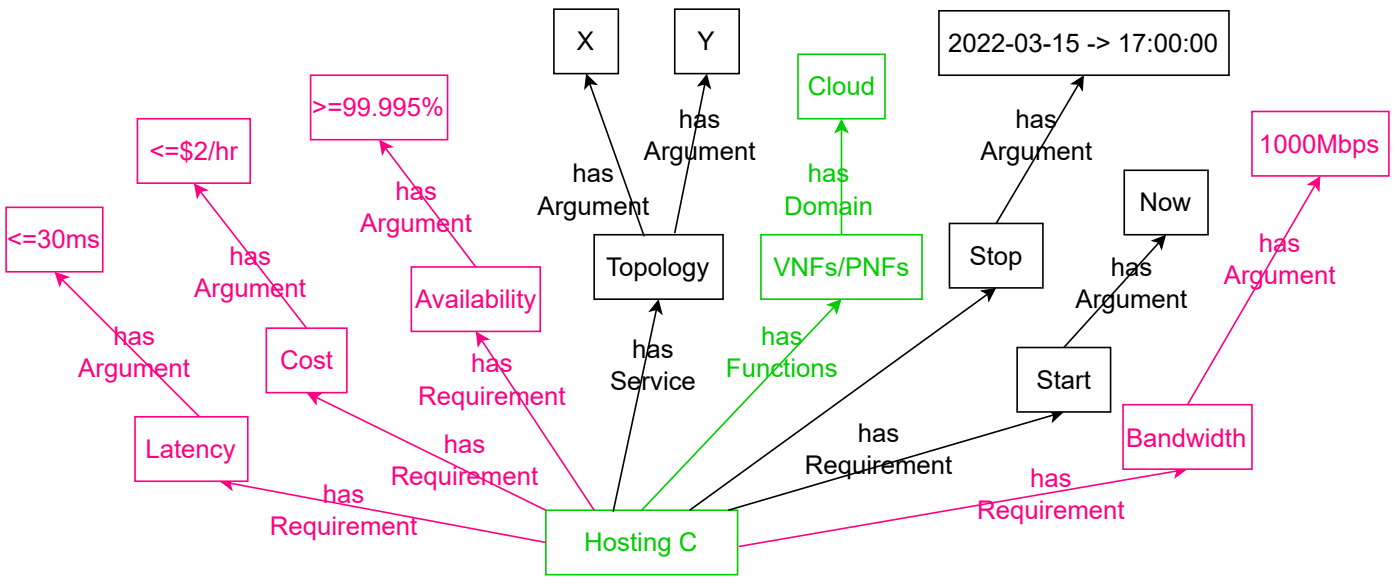

Fig. 15. Sub-intent obtained after Non-Functional Decomposition of intent to host feature C on Cloud.

*5.1.5 Intent Processing Methods:* To process an intent from its specified form to a well defined RMSO interpretable format, five main intent processing methods are: (1) Natural Language Processing (NLP), (2) Deterministic Finite Automata (DFA), (3) Context Free Grammars (CFG), (4) Semantic Graphs and (5) Genetic Programming. All these methods are found to be used independently as well as in conjunction with each other. NLP (also known as computational linguistics [111]) is a method to interpret and manipulate human language to a computer native language and is used to process intents specified in CNL or other informal languages [4, 55, 127, 135]. Modern practitioners and researchers have started to use machine learning (ML) methods to improve the efficiency and effectiveness of classic NLP methods as a result of advancements in big data methods for managing and analyzing large amounts of data. Chao et al. [19], Jacobs et al. [47], Angi et al. [7], Yang et al. [131] and Souihi et al. [108] used NLP in conjunction with ML to process the intents. DFA is a finite state machine which takes strings as input and perform actions and produces an output for each state transition. DFA machines are used to extract the strings/keywords of interest from a high-level intent and convert them to machine compatible values. Scheid et al. [97], Yang et al. [132] and Kim et al. [54] used DFA to extract strings, for example, names of the users from high-level intents and replace them with corresponding IP addresses by using database maintaining IP addresses of all users. CFG is a formal grammar with certain types of production rules required to process an intent to get details about system design and configuration parameters [28]. Most of the solutions use CFG in alliance with DFA to process an intent [21, 24, 97, 106]. Semantic graph is a network with labeled edges and nodes used to represent semantic relationships between concepts [109]. These are very useful to maintain the knowledge bases required to process an intent and can be used either independently [2, 3, 6, 32, 50, 61] or in association with other methods, i.e., ML [98]. Hireche et al. [45] used intent processing methods based on Genetic programming.

*5.1.6 Configuration Languages:* After processing an intent, output is generated in languages called 'Configuration Languages'. Based on the abstraction level of the configuration language, we have divided them into two categories: (1) Low-Level and (2) High-Level configuration languages.

- *Low-Level Languages:* An intent processed into a low-level language has no abstraction from the language acceptable by RMSO responsible of service deployment. No intermediate process is required to convert the output generated in a low-level language to RMSO acceptable language. It can be directly accepted as input by the underlying system. OASIS TOSCA [52, 61] and Datalog [50] are the low-level languages in which the service design solutions are generated and applied directly to the underlying RMSO.
- *High-Level Languages:* An intent processed into a high-level language has a high-level of abstraction from the languages acceptable by RMSOs. The intent processing outputs generated in high-level languages need an intermediate processing unit (some kind of compiler or interpreter) to make them acceptable by RMSO for service deployment. JSON [24, 52, 59, 98], YAML [15, 19, 52, 127], Yet Another Next Generation (YANG) [112], Nile [47], Multi Resource Markup Language (MRML)[78], RDF graphs [20, 42, 55, 66], Unified Modeling Language (UML) models [81], P4 [45, 91, 92] and proprietary configuration languages [64, 97, 110, 117] are the high-level languages to represent output of an IH.

### 5.2 Autonomous Deployment and Orchestration

This activity deals with hosting the decomposed intents (system design and configuration parameters) on the underlying virtual and physical infrastructure. An intelligent resource manager and service orchestrator (RMSO) accepts the generated service design and configuration information. It performs the required changes in the underlying infrastructure by provisioning and allocating the virtual/physical resources across multiple domains/sub-systems to fulfill an intent. Figure 16 provides the taxonomy for autonomous deployment and orchestration representing SLA parameters the intent stakeholders (users and service providers) target and various resource provisioning and management methods to host and fulfill the intents. Table 7 summarizes the existing research works covering autonomous deployment and orchestration of intents.

*5.2.1 SLA Parameters:* Service Level Agreement (SLA) is a contractual agreement between two parties, i.e., service provider and its consumer written in a legal format which both parties are abide to follow during the specified period of the contract. Specification of an SLA is usually done in the measurable terms representing what a service provider will furnish in terms of QoS parameters, a.k.a SLA parameters. Additionally, it covers the penalties the service provider will pay, for example, monetary compensation when the promised service is not maintained or delivered. It is also possible that two or more parties come together to provide a service, which is a case in IDSM systems where edge, communication and cloud service providers create an ecosystem for providing a service (Figure 3). In such cases, an SLA will be a multi-party SLA with domain/sub-system specific SLA parameters. Based on the characteristics of SLA parameters targeted by the intents, we have divided them in two categories: (1) Networking and (2) Computing SLA parameters.

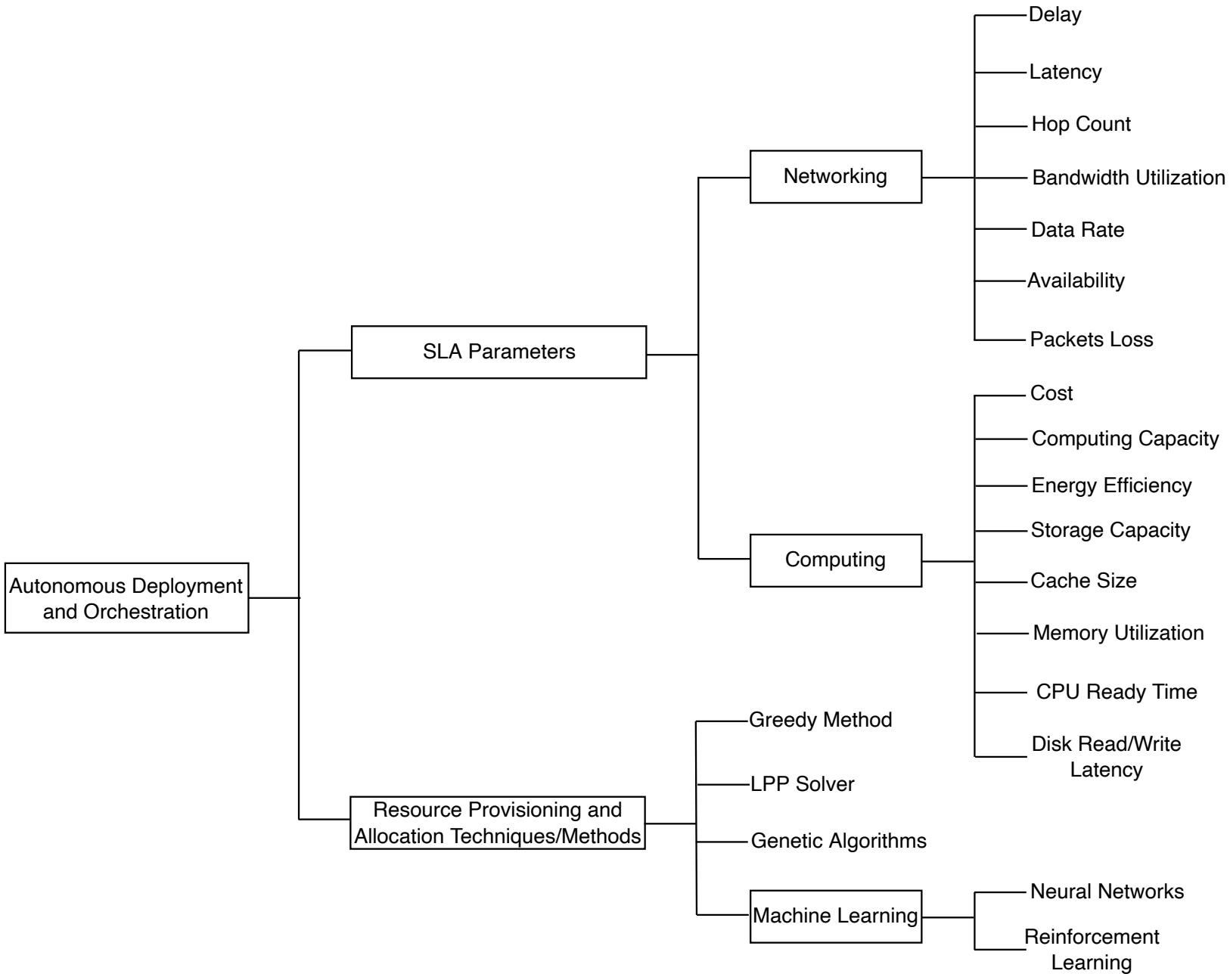

Fig. 16. Taxonomy for Autonomous Deployment and Orchestration Activity

Table 7. Summary of existing works considering Autonomous Deployment & Orchestration Activity Taxonomy

| Reference | Autonomous Deployment & Orchestration | | Reference | Autonomous Deployment & Orchestration | |
|---|---|---|---|---|---|
| | SLA Parameters | Intent Realizing Resource Allocation Techniques/Methods | | SLA Parameters | Intent Realizing Resource Allocation Techniques/Methods |
| [2] | BU | GM, LPP | [35] | C, SC, CS | GM, GA |
| [121] | BU | GM | [59] | HC | LPP |
| [80] | HP | GM | [105] | D, C | RL |
| [67] | BU, L, A | GM | [88] | T, HC, BU | GM |
| [4] | DR | NG | [41] | HC, BU | NG |
| [71] | EE, CU | NG | [131] | D, BU, L | NN, RL |
| [73] | HC, BU, L, A | NG | [53] | BU, L | NG |
| [48] | BU | NG | [44] | D, BU, L, A, CU, MU, SC | GM |
| [8] | DR, L | NG | [94] | D, BU, L | GM |
| [10] | D, L, CU | GM | [63] | D, HC, BU | GM |
| [135] | CU, MU, SC | GM | [18] | D, BU, PL | GM |
| [126] | CRT, DRWL | NG | | | |

**NG:** Not given, **D:** Delay, **L:** Latency, **HC:** Hop count, **BU:** Bandwidth utilization, **DR:** Data rate, **A:** Availability, **PL:** Packets loss, **C:** Cost
**CU:** Computing utilization, **EE:** Energy efficiency, **SC:** Storage capacity, **CS:** Cache size, **MU:** Memory utilization, **CRT:** CPU ready time
**DRWL:** Data read/write latency, **GM:** Greedy method, **LPP:** LPP solver, **GA:** Genetic algorithms, **NN:** Neural networks, **RL:** Reinforcement learning

- *Networking SLA parameters:* SLA parameters for networks are the performance parameters within which a network service is required to be provided to fulfill an intent. Various networking SLA parameters that are targeted by intent stakeholders include delay, latency, hop count, bandwidth utilization, data rate, availability and packet loss (Figure 16). Delay (refers to transmission delay) describes the time required to transmit/transport a data packet from one end (source) of the network to the other (destination) [18, 105, 131]. Latency is also a measure of delay representing a round-trip time taken by a data packet to reach its destination and back again [10]. Hop count refers to the number of devices/nodes, usually routers, that a data packet passes through from its source to destination. Hop count is used by intents with environment access/visibility non-functional attributes requesting for a security as a service. Kumar et al. [59] specified an intent with hop-count as an SLA parameter with an objective to reduce the number of hops to minimize the cost of security rules placement. Bandwidth utilization specified in an intent as a service requirement refers to the maximum data transfer rate required over a specific connection [2, 52, 67, 88, 121]. As SLA parameters, all these metrics are specified by using their upper bound values. If the observed value of any of these parameters is more than the specified value then SLA violation occurs which leads to an intent being unsatisfied.

Data rate denotes the transmission speed, or the number of bits per second required to transfer to fulfill an intent [4, 8, 88]. Availability of a network is a critical SLA parameter which represents the level of accessibility, connectivity and performance of a network in terms of its uptime (network is fully operational) over a specific time interval [44, 67]. Data rate and availability are specified in an intent differently from other networking SLA parameters discussed above. When the obtained data rate or availability is below the intended values, it is considered as intent violation.

- *Computing SLA parameters:* These parameters target the performance of computing and storage infrastructure provisioned and allocated in domains/sub-systems (edge, CSP and cloud) selected to satisfy an intent. The parameters of interest are; cost, computing capacity, energy efficiency, storage capacity, cache size, memory utilization, CPU ready time and Disk read/write latency (Figure 16). The Cost of the service is one of the parameters both service users and providers are interested in the most to regulate. Besides cost, other computing SLA parameters used to specify the intents are computing capacity (CPU count, its utilization and cache size) and energy consumption of the computing infrastructure provisioned to fulfill an intent. Mehmood et al. [71] proposed a method to regulate the CPU utilization and energy efficiency of the computing infrastructure to meet the profit goals for both service users and providers while fulfilling the intents. Elhabbash et al. [35] exploited storage capacity and cache size as internal SLA parameters to satisfy an intent of a user with minimum cost.

*5.2.2 Resource Provisioning and Allocation Techniques/Methods:* To satisfy SLA parameters in intents, RMSOs perform provisioning and allocation of virtual/physical resources across multiple domains/sub-systems identified during the intent decomposition. In this section, such resource provisioning and allocation techniques used to fulfill the intents are discussed.

- *Greedy Method:* It is a simple and intuitive method to design algorithms which makes local optimal choice at each step to obtain an approximate global optimal solution. In crux, it constructs the optimal solution piece by piece. Resource management solutions for IDSM systems based on greedy method choose the best physical/virtual resources available at the moment to host a service request. The solution then extends iteratively to other service request instances to achieve a global optimal solution. Abhashkumar et al [2], Elhabbash et al. [35] and He et al. [44] used greedy method based algorithms for resource management and allocation to fulfill the intents.
- *Linear Programming Problem Solver:* Linear programming (LP) is a mathematical optimization technique to determine the optimal allocation of scarce resources with having linear objective functions and relations among the variables corresponding to resources. Kumar et al. [59] formulated and solved the problem of traffic blocking rule placement by using LP with minimum cost while satisfying the security requirements specified as an intent.

- *Genetic Algorithms:* It is a search-based technique inspired from the process of biological evolution and can be used for solving resource optimization problems with linear or non-linear and continuous or non-continuous objective functions. Elhabbash et al. [35] used genetic algorithm based approach to maximize the number of intents being served with optimal selection of services offered by the service provider.
- *Machine Learning:* Resource management methods employing data analytics and model building are covered in this type. Neural Networks and Reinforcement learning are the two commonly used ML methods. Yang et al. [131] used a reinforcement learning based deep Q network (DQN) method for resource composition satisfying the requirements of an intent.

## 5.3 Monitoring and Awareness

The primary task of monitoring and awareness activity is to provide periodic feedback to intent stakeholders about the status of the intents. This activity also identifies and predicts the anomalies (outage/failure or congestion/resource over utilization) in the system that can impact the fulfillment of the intents. Intent-driven service management (IDSM) system performs periodic data collection from the physical and virtual resources. It uses the data to perform the analytical operations to evaluate the current state of the system. Obtained results are used to determine if the current performance of the system is fulfilling the intents and able to host new intents. If the telemetry results are found to be satisfactory w.r.t the hosted intent service level agreement (SLA) parameters (Section 5.2.1), the existing resource management policy remains unchanged. Otherwise, refinement/remediation activities (Section 5.4) gets activated, autonomously to fix the system's performance and avoid any anomaly which can impact the fulfillment of the intents. Figure 17 provides the taxonomy for monitoring and awareness activity representing various performance monitoring and prediction methods, key performance indicators (KPIs) and performance challenges. Table 8 summarizes the existing research works covering monitoring and awareness activity.

*5.3.1 Performance Challenges:* During intent's fulfillment, IDSM systems face various performance challenges, such as resource failures, network congestion, resource overloading and resource scalability. Regular monitoring of KPIs (Section 5.3.2) is required to avoid/handle the occurrence of events posing such challenges.

Table 8. Summary of existing works considering Monitoring and Awareness Activity Taxonomy

| Reference | Monitoring & Awareness | | | | Reference | Monitoring & Awareness | | | |
|---|---|---|---|---|---|---|---|---|---|
| | Performance Challenges | Performance Monitoring Methods | Key Performance Indicators | Performance Prediction Methods | | Performance Challenges | Performance Monitoring Methods | Key Performance Indicators | Performance Prediction Methods |
| [110] | RF, RO | AM, PM | CU, MU, LU | SP | [117] | NC, RO | AM | AB | SP |
| [95] | RF, RO | AM | LU | TSA | [30] | RF, NC | AM | PD, L | SP |
| [96] | RO | AM | AB, PD | SP | [3] | RO, RS | AM, PM | CU, MU, AB | SP |
| [52] | RO | AM | AB | NN | [131] | RF | AM | AB, PD | NN |
| [128] | RF | AM, PM | CU | LR | [53] | NC, RO | AM | AB, PD, LU | NN |
| [137] | RO | PM | CU | NN | [33] | NC | AM | LU | SP |
| [1] | RO | PM | CU, MU, S, T | NN | [31] | NC | NG | T, PD, J | SP |
| [25] | RO | PM | CU, MU | NN | [40] | NG | AM, PM | T | NG |
| [120] | NC, RO | PM | L | NG | [93] | NC, RO | PM | CU | KC, KNC |
| [8] | RO, RS | PM | AB, T, PD, L | SP | [10] | RO | PM | CU | SP |
| [45] | NC | AM, PM | T | NN | [138] | RS | AM | CU, MU, S | SP |
| [68] | TC, RO | PM | T, PRT | SP | [126] | RO, RS | PM | CU, S, L | NN |

**NG:** Not given, **RF:** Resource failures, **NC:** Network congestion, **RO:** Resource overloading, **RS:** Resource scalability, **CU:** CPU usage, **MU:** Memory usage
**S:** Storage, **AB:** Achieved bandwidth **PD:** Packets dropped, **PRT:** Packets received/transmitted, **L:** Latency, **LU:** Link utilization, **T:** Throughput, **J:** Jitter
**HC:** Hop count, **AM:** Active monitoring, **PM:** Passive monitoring, **SP:** Static prediction, **NN:** Neural networks, **TSA:** Time series analysis, **LR:** Linear regression
**KC:** K-means clustering, **KNC:** K-nearest neighbours classification

- *Resource Failures:* Occurrence of failures is inevitable and a biggest challenge that all the systems face, including IDSM systems. There are various reasons that can cause the failure of resources (both physical and virtual) and consequently causes the service outage [104].

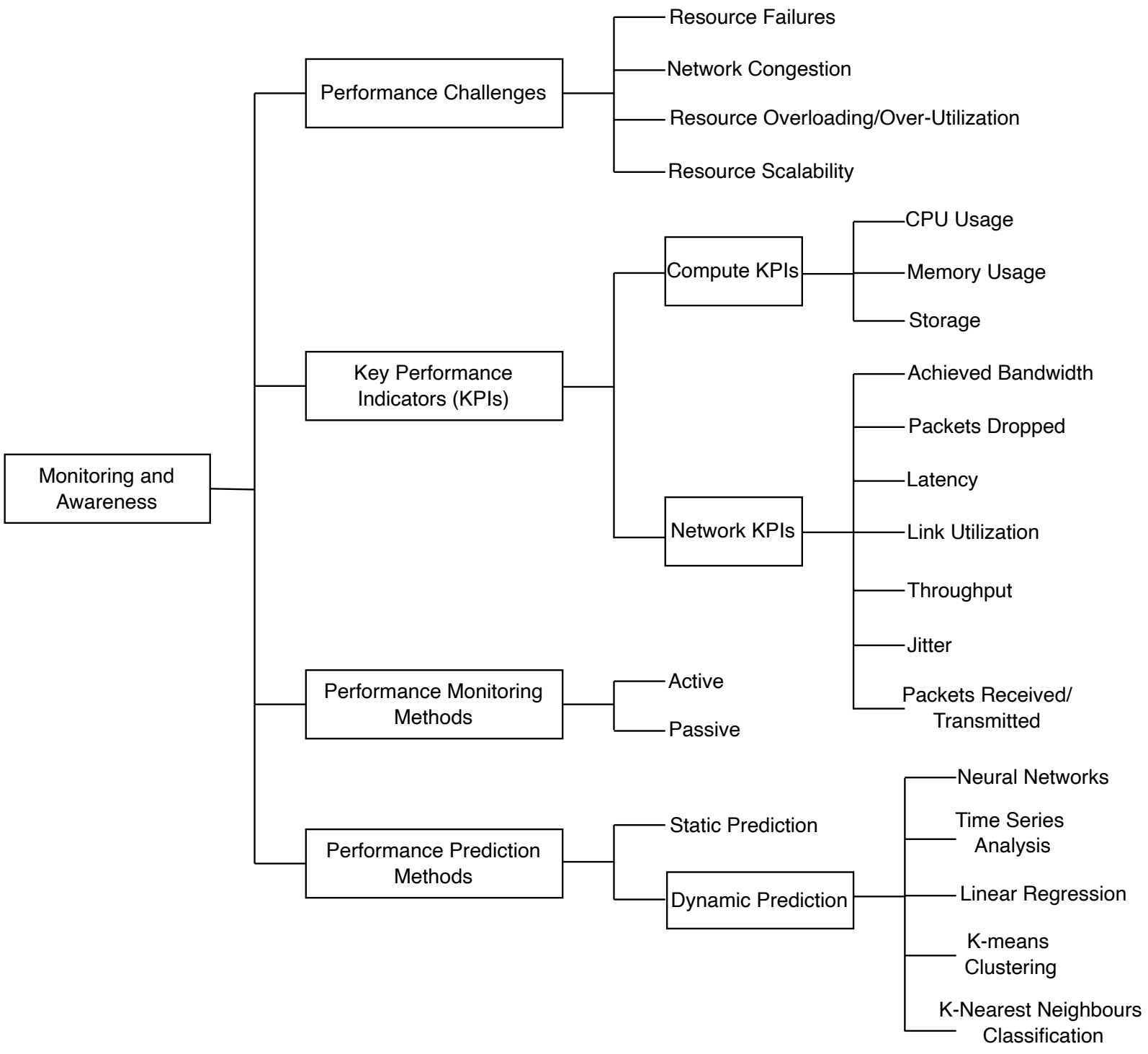

Fig. 17. Taxonomy for Monitoring and Awareness Activity

Identical reasons are found for failures in IDSM systems. Sung et al. [110] identified database application and replicated service failures as the cause of service outage. Sanvito et al. [95], Davoli et al. [30], Yang et al. [131], Wu et al. [128] considered link failures and computing resource failures impacting the service connectivity in their IDSM solutions.

- *Network Congestion:* A spike in the demand of a service increases the data transmission/traffic. This can exceed the capacity of the network and may lead to the network congestion. Consequently, it impacts the quality of a service and can cause a service outage or makes it inaccessible [39]. Tsuzaki et al. [117], Hireche et al. [45], Martini et al. [68] considered network congestion as a performance challenge in their intent management solutions.
- *Resource Overloading/Over-utilization:* Over utilization of provisioned computing resources, such as CPU, memory (both RAM and cache) and storage, can also cause the performance degradation in IDSM systems. This consequently impacts the fulfillment of the intents. Saraiva et al. [96], Aklamanu et al . [3], Khan et al. [52], Abbas et al. [1], Ustok et al. [120] proposed IDSM solutions dealing with the challenge of resource overloading.
- *Resource Scalability:* It is the ability of a service management system to provision the resources, autonomously to handle the workload growth. However, performing resource scalability in IDSM systems without impacting the intents and; increasing the cost and operational complexity is a challenge. Aklamanu et al . [3], Baktir et al. [8], Zheng et al. [138] addressed the challenge of resource scalability in the proposed IDSM solutions.

*5.3.2 Key Performance Indicators (KPIs):* To get the quantifiable measurements required to gauge the compliance of SLA parameters, KPIs play a significant role [103]. Collecting, processing and analyzing the data for KPIs of interest provides insight into the system's performance. The obtained information is further used to compare against the SLA parameters. It is to measure the satisfaction level of intents and to identify or predict any performance challenges. In case, a performance diversion is found or predicted to happen, IDSM system takes performance corrective decisions (Section 5.4.1), autonomously. Based on the characteristics of the components (both virtual and physical) involved in serving the intents, we have divided the KPIs in following two classes:

- *Compute KPIs:* These KPIs are used to measure the utilization level of provisioned computing resources, such as CPU, memory and storage. Researchers are mainly focused on CPU and memory usage to improve intent satisfaction levels. Aklamanu et al. [3], Collet et al. [25] used CPU and memory usage KPI values whereas Abbas et al. [1], Zheng et al. [138], Wu et al. [126] used storage KPIs as well to handle the challenges of resource overloading and failures.
- *Network KPIs:* These KPIs are essential to determine the performance of networking components (both physical and virtual) required to fulfill the intents. Achieved bandwidth [52, 96, 117], packets dropped [31, 131], latency [8, 30, 120], link utilization [53, 95], throughput [1], jitter [31] and packets received/transmitted [68] are the network KPIs the researchers are using to evaluate the performance of their IDSM solutions.

*5.3.3 Performance Monitoring Methods:* Two types of monitoring methods are used to monitor the KPIs representing the performance of IDSM systems: (1) Active and (2) Passive Monitoring.

- *Active Monitoring:* This method is also known as synthetic monitoring. It injects the test traffic (synthetic traffic) into the system to get the real-time view of its performance. Khan et al. [52], Dzeparoska et al. [33], Ustok et al. [120] are some of the works using active monitoring method to monitor and analyze the fulfillment of the intents.
- *Passive Monitoring:* This method involves capturing and analyzing the real traffic flow, periodically representing the performance of serving components of the system. Sung et al. [110], Yang et al. [131], Wu et al. [128], Zheng et al. [137], Abbas et al. [1] employed passive monitoring to observe the parameters of interest in their proposed IDSM solutions.

*5.3.4 Performance Prediction Methods:* To fulfill the intents, a reliable prediction of service performance or an event that can affect the performance is critical. Furthermore, having efficient and accurate performance prediction methods provide a leverage to the service providers during intent negotiation. It helps to advise the users about the possible service performance degradation if they choose not to select the alternative solutions provided by the service provider (Section 5.1.2). This facilitates both the service users and providers to draft the rich and accurate SLAs and avoid any legal conflicts that can occur because of SLA violations. Performance prediction methods use the KPI values to predict service performance challenges (Section 5.3.1) that can impact the fulfillment of intents. We have divided the performance prediction methods in following two classes:

- *Static Prediction:* In the Static Prediction methods, the occurrence of an event is predicted based on a static threshold value for a variable which remains unchanged until the manual changes are made. The threshold values are obtained and set by the system administrators based on their experience from previous runs. For example, if the system outage is happening at the certain utilization level of a CPU then it will be marked as a threshold value for CPU utilization. When the KPIs (both for compute and network components) under observation reaches the threshold values, performance corrective methods are triggered, autonomously to safeguard the intents. Static prediction methods are the most commonly used methods because of the simplicity of their application. Sung et al. [110], Aklamanu et al. [3], Davoli et al. [30], Saraiva et al. [96], Tsuzaki et al. [117], Dzeparoska et al. [33], de Sousa et al. [31], Martini et al. [68] used static threshold values to predict the service performance challenges.

- *Dynamic Prediction:* The drawback of the static prediction methods is that they do not evolve with time, such that the threshold values remain same until they are changed manually. However, due to the autonomous nature of IDSM systems, employment of static prediction methods is not an optimal solution. As an alternate, dynamic performance prediction methods is the solution of choice where the threshold values change with time in an autonomous manner by employing ML based methods. Yang et al. [131], Khan et al. [52], Zheng et al. [137], Abbas et al. [1], Hireche et al. [45] used Neural networks for performance prediction. Sanvito et al. [95], Wu et al. [128], Rivera et al. [93] used time series analysis, linear regression and K-mean clustering and classification enabled dynamic performance prediction methods, respectively in their proposed IDSM solutions.

## 5.4 Dynamic Optimization and Remediation

Based on the telemetry results, the intent-driven service management (IDSM) systems autonomically optimize their performance to meet the SLA parameters required to fulfill the intents. Performance optimization includes internal reconfiguration of computing and networking resources to safeguard the intents from any predicted anomaly or increase the overall efficiency of the system (Section 5.4.2). Figure 18 provides the taxonomy for dynamic optimization and remediation activity classifying the methods for intent guarantee management and performance optimization and remediation. Table 9 summarizes the existing research works covering dynamic optimization and remediation activity.

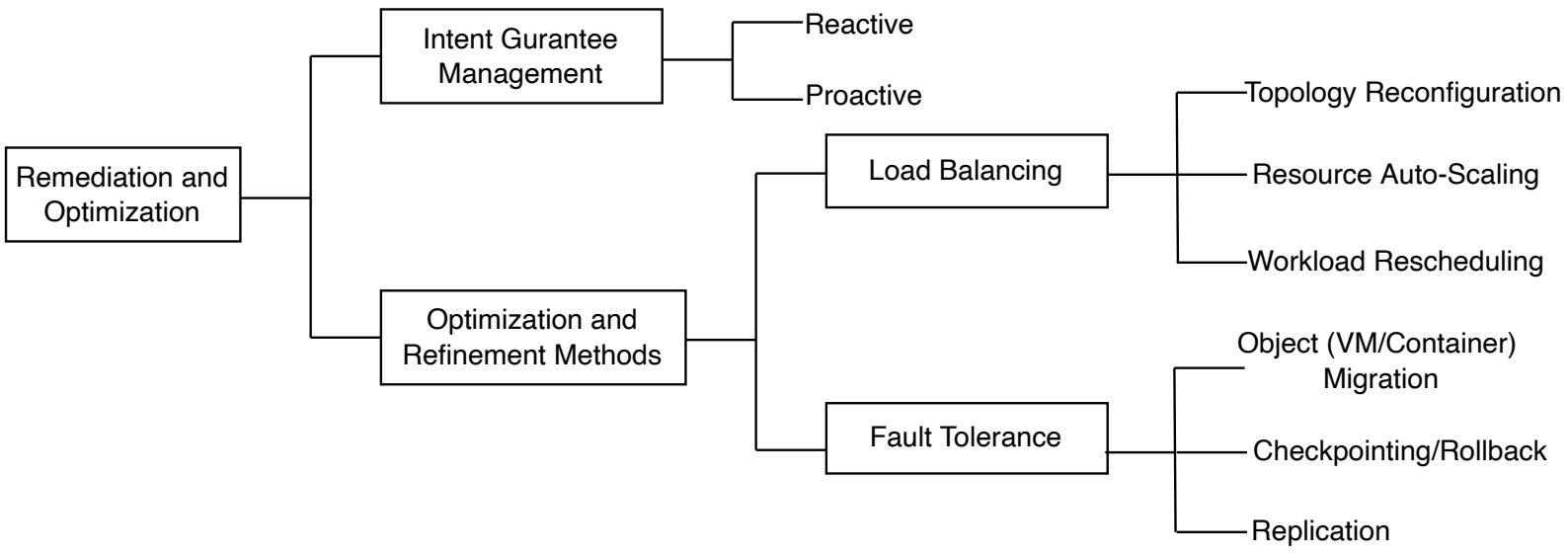

Fig. 18. Taxonomy for Dynamic Optimization and Remediation Activity

Table 9. Summary of existing works considering Dynamic Optimization and Remediation Activity Taxonomy

| Reference | Dynamic Optimization & Remediation | | Reference | Dynamic Optimization & Remediation | |
|---|---|---|---|---|---|
| | Intention Guarantee Management | Optimization & Refinement Methods | | Intention Guarantee Management | Optimization & Refinement Methods |
| [110] | R | C/R | [117] | R | TR |
| [95] | P | TR | [30] | P | RA, Rep. |
| [96] | R | TR | [52] | P | TR |
| [131] | R | TR, OM | [33] | R | TR |
| [1] | P | RA | [93] | R | TR |
| [8] | P | RA, WR, OM | [10] | R | OM |
| [45] | R | TR | [138] | R | WR |
| [68] | R | TR | | | |

**R:** Reactive, **P:** Proactive, **TR:** Topology reconfiguration, **RA:** Resource auto-scaling, **WR:** Workload Rescheduling, **OM:** Object migration
**C/R:** Checkpointing/rollback, **Rep.:** Replication

*5.4.1 Intent Guarantee Management:* Methods to guarantee the fulfillment of intents are divided into two categories: (1) Reactive and (2) Proactive.

- *Reactive Management:* In this method, measures are taken after the occurrence of an event. For example, in case of checkpointing used as a fault tolerance method, recovery takes place from the last saved checkpoint after the occurrence of a failure event [101]. Sung et al. [110]

and Yang et al. [131] used reactive methods for failure management where as Tsuzaki et al. [117], Saraiva et al. [96], Dzeparoska [33], Hireche et al. [45], Zheng et al. [138] used reactive methods to optimize the performance of IDSM systems.
- *Proactive Management:* In this method, measures are taken before the occurrence of an event. These methods are prediction driven methods where the occurrence of an event is predicted by using machine learning (ML) and data analytic operations. The efficacy of the proactive management methods depend upon the accuracy of prediction algorithms. Sanvito et al. [95], Davali et al. [30] and Baktir et al. [8] used proactive methods to provide fault tolerance in IDSM systems to safeguard the intents from the failures. Khan et al. [52] and Abbas et al. [1] used the proactive management methods to optimize the performance of IDSM systems.

*5.4.2 Optimization and Remediation Methods:* The methods used to guarantee the optimal fulfillment of intents are divided into two categories: (1) Load balancing and (2) Fault tolerance. The load balancing methods are typically used for performance optimization of IDSM systems to make them more efficient. Whereas, fault tolerance methods are employed to safeguard the intents against the failures in IDSM systems. The details of both categories are as follows:

- *Load Balancing:* It is the performance optimization method used to increase the efficiency of the virtual/physical resources provisioned to host the services fulfilling the intents. This is to avoid a service breakdown or periodically optimize the efficiency of the system in terms of energy consumption [36], bandwidth utilization [79] and many more. Topology reconfiguration, resource auto-scaling and workload rescheduling are the load balancing mechanisms which can be triggered reactively or proactively. Tsuzaki et al. [117], Saraiva et al. [96], Hireche et al. [45] triggered the topology reconfiguration by re-routing the traffic reactively if the bandwidth usage and throughput of a link exceeded a predefined threshold value. Davoli et al. [30], Abbas et al. [1] employed auto-scaling by adding extra resources proactively to avoid any resource scarcity. Zheng et al. [138] applied the rescheduling of intent requests according to their temporal (start and stop timestamps) and spacial attributes (targeting similar physical and/or virtual components) to resolve the conflicts, autonomously.
- *Fault Tolerance:* To guarantee the fulfillment of intents, IDSM systems need to manage the service failures. Various mechanisms are used to provide fault tolerance in IDSM systems, such as object (VM or container) migration, checkpointing/rollback and replication. Davoli et al. [30] maintained replicated copy of each transmitted packet to recover from, in case a transmitted packet is lost. Sung et al. [110] used the periodical checkpointing to save the healthy state of the IDSM system and used it to recover from the failures. Yang et al. [131] used object migration to migrate the workload from a predicted to be failed computing resource to a healthy one to safeguard the intents from failures.

## 6 DISCUSSION

This section discusses the principal findings of our systematic review. The discussion covers the critical analysis of all the considered works and highlights the key observations. It also highlights the open challenges and future research directions in service level agreement (SLA) management in intent-driven service management (IDSM) systems.

### 6.1 Critical Analysis and Key Observations

All studies considered in the survey are critically analyzed and compared in Table 10. The analysis drove the key observations made on the basis of the IDSM activities covered in a solution, its scale (multi-domain or single-domain), area of focus and employment of machine learning (ML) methods. All the observations are supported by the quantitative analysis represented in Figure 19.

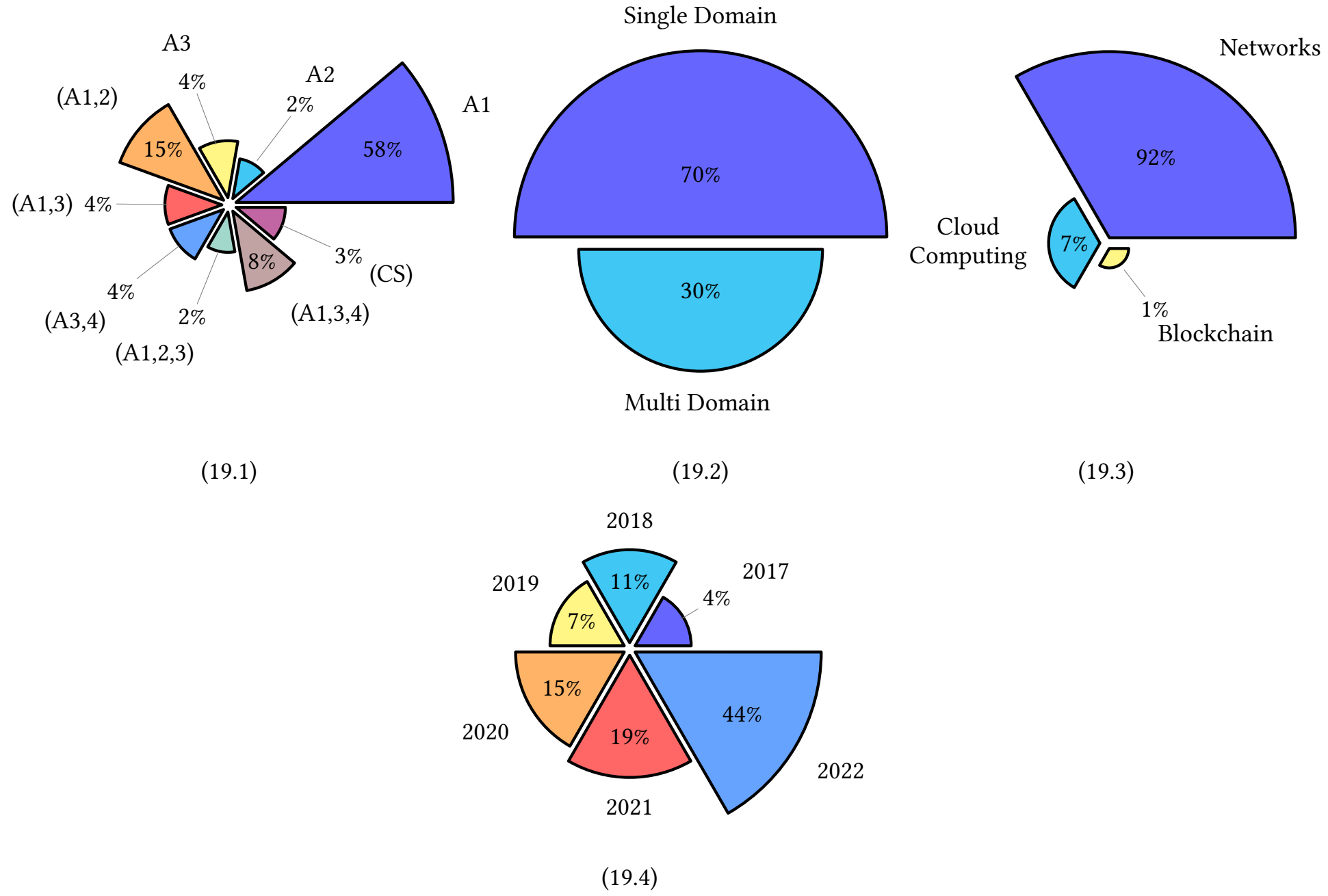

Fig. 19. Quantitative analysis representing the strengths and weaknesses of the current state-of-the-art. Figure shows the distribution of research articles according to the (1) covered IDSM activities (2) scale of the solution (3) area of application (4) use of machine learning.

Table 10. Analysis of the research articles, with highlights of their strengths and weaknesses.

| Authors | Year | IDSM Activities | | | | Scale of Solution | | Area of Focus | | | Use of ML |
|---|---|---|---|---|---|---|---|---|---|---|---|
| | | A1 | A2 | A3 | A4 | SD | MD | Networks | Cloud computing | Blockchain | |
| Sung et al.[110] | 2016 | ✓ | | ✓ | ✓ | ✓ | | ✓ | | | |
| Scheid et al.[97] | 2017 | ✓ | | | | ✓ | | ✓ | | | ✓ |
| Tsuzaki et al.[117] | 2017 | ✓ | | ✓ | ✓ | | ✓ | ✓ | | | |
| Abhashkumar et al.[2] | 2017 | ✓ | ✓ | | | ✓ | | ✓ | | | |
| Kang et al.[50] | 2017 | ✓ | | | | ✓ | | | ✓ | | |
| Alsudais et al.[6] | 2017 | ✓ | | | | ✓ | | ✓ | | | |
| Sköldström et al.[106] | 2017 | ✓ | | | | ✓ | | ✓ | | | |
| Liu et al.[65] | 2018 | ✓ | | | | | ✓ | ✓ | | | ✓ |
| Comer et al.[26] | 2018 | ✓ | | | | | ✓ | ✓ | | | |
| Sanvito et al.[95] | 2018 | | | ✓ | ✓ | ✓ | | ✓ | | | ✓ |
| Yang et al.[132] | 2018 | ✓ | | | | ✓ | | ✓ | | | |
| Elhabbash et al.[35] | 2018 | ✓ | ✓ | | | ✓ | | ✓ | | | |
| Dzeparoska et al.[32] | 2018 | ✓ | | | | | ✓ | ✓ | | | |
| Vilalta et al.[121] | 2018 | ✓ | ✓ | | | | ✓ | | ✓ | | |
| Tuncer et al.[118] | 2018 | ✓ | | | | ✓ | | ✓ | | | |
| Esposito et al.[37] | 2018 | ✓ | | | | ✓ | | ✓ | | | |
| Chao et al.[19] | 2018 | ✓ | | | | ✓ | | | ✓ | | ✓ |
| Monga et al.[78] | 2018 | ✓ | | | | | ✓ | ✓ | | | |
| Kiran et al.[55] | 2018 | ✓ | | | | | ✓ | ✓ | | | |
| Davoli et al.[30] | 2018 | ✓ | | ✓ | ✓ | | ✓ | ✓ | | | |
| Szyrkowiec et al.[112] | 2018 | ✓ | | | | ✓ | | ✓ | | | |
| Wang et al.[122] | 2019 | ✓ | | ✓ | | ✓ | | ✓ | | | |
| Saraiya et al.[96] | 2019 | | | ✓ | ✓ | ✓ | | ✓ | | | |
| Riftadi et al.[92] | 2019 | ✓ | | | | ✓ | | ✓ | | | ✓ |
| Wu et al.[127] | 2019 | ✓ | | | | ✓ | | | ✓ | | |
| Riftadi et al.[91] | 2019 | ✓ | | | | ✓ | | ✓ | | | |
| Aklamanu et al.[3] | 2019 | ✓ | | ✓ | | | ✓ | ✓ | | | |
| Borsatti et al.[15] | 2019 | ✓ | | | | ✓ | | ✓ | | | |
| Tian et al.[114] | 2019 | ✓ | | | | | ✓ | ✓ | | | |
| Kumar et al.[59] | 2019 | ✓ | ✓ | | | ✓ | | ✓ | | | |
| Chen et al.[20] | 2019 | ✓ | | | | ✓ | | ✓ | | | |
| Chung et al.[24] | 2019 | ✓ | | | | | ✓ | ✓ | | | |
| Jacobs et al.[47] | 2019 | ✓ | | | | ✓ | | ✓ | | | ✓ |
| Scheid et al.[98] | 2020 | ✓ | | | | ✓ | | | | ✓ | |
| Khan et al.[52] | 2020 | ✓ | | ✓ | ✓ | | ✓ | ✓ | | | ✓ |
| Chung et al.[23] | 2020 | ✓ | | | | ✓ | | ✓ | | | |
| Ujcich et al.[119] | 2020 | ✓ | | | | ✓ | | ✓ | | | |
| Alalmaei et al.[4] | 2020 | ✓ | | | | ✓ | | | ✓ | | |
| Mahtout et al.[66] | 2020 | ✓ | | | | | ✓ | ✓ | | | ✓ |
| Nagendra et al.[80] | 2020 | ✓ | ✓ | | | ✓ | | ✓ | | | |
| Gao et al.[41] | 2020 | ✓ | | | | ✓ | | ✓ | | | |
| Shi et al.[105] | 2020 | ✓ | ✓ | | | | ✓ | ✓ | | | ✓ |
| Ribeiro et al.[89] | 2020 | ✓ | | | | ✓ | | ✓ | | | |
| Nazarzadeoghaz et al.[81] | 2020 | ✓ | | | | ✓ | | ✓ | | | |
| Kim et al.[54] | 2020 | ✓ | | | | | ✓ | | ✓ | | |
| Wang et al.[124] | 2020 | ✓ | | | | ✓ | | ✓ | | | |
| Marsico et al.[67] | 2020 | ✓ | ✓ | | | ✓ | | ✓ | | | |
| Rafiq et al.[88] | 2020 | ✓ | ✓ | | | ✓ | | ✓ | | | |
| Mehmood et al.[71] | 2020 | ✓ | ✓ | | | | ✓ | ✓ | | | |
| Yang et al.[131] | 2020 | ✓ | ✓ | ✓ | ✓ | | ✓ | ✓ | | | ✓ |
| Zhang et al. [134] | 2021 | ✓ | | | | ✓ | | ✓ | | | |
| Wu et al. [128] | 2021 | | | ✓ | | ✓ | | ✓ | | | ✓ |
| Gritli et al. [43] | 2021 | ✓ | | | | | ✓ | ✓ | | | |
| Mehmood et al. [73] | 2021 | ✓ | ✓ | | | ✓ | | ✓ | | | |
| Khan et al. [53] | 2021 | ✓ | ✓ | ✓ | | ✓ | | ✓ | | | ✓ |
| Mercian et al. [76] | 2021 | ✓ | | | | ✓ | | ✓ | | | |
| Zheng et al. [137] | 2021 | | | | ✓ | ✓ | | ✓ | | | ✓ |
| Bensalem et al. [11] | 2021 | ✓ | | | | ✓ | | ✓ | | | |
| Bezahaf et al. [12] | 2021 | ✓ | | | | ✓ | | ✓ | | | |
| Ouyang et al. [85] | 2021 | ✓ | | | | ✓ | | ✓ | | | |
| Dzeparoska et al. [33] | 2021 | ✓ | | ✓ | ✓ | | ✓ | ✓ | | | |
| Abbas et al. [1] | 2021 | ✓ | | ✓ | ✓ | | ✓ | ✓ | | | ✓ |
| el houda Nouar et al. [34] | 2021 | ✓ | | | | ✓ | | ✓ | | | |
| Kuwahara et al. [61] | 2021 | ✓ | | | | ✓ | | ✓ | | | |
| de Sousa et al. [31] | 2021 | ✓ | | ✓ | | | ✓ | ✓ | | | |
| Jacobs et al. [48] | 2021 | ✓ | ✓ | | | ✓ | | ✓ | | | ✓ |
| Gomes et al. [42] | 2021 | ✓ | | | | | ✓ | ✓ | | | |
| Curtis-Black et al. [29] | 2021 | ✓ | | | | | ✓ | ✓ | | | |
| Collet et al. [25] | 2022 | | | ✓ | | | ✓ | ✓ | | | ✓ |
| McNamara et al. [70] | 2022 | ✓ | | | | ✓ | | ✓ | | | |
| He et al. [44] | 2022 | ✓ | ✓ | | | ✓ | | ✓ | | | |
| Fernández et al. [40] | 2022 | ✓ | | ✓ | | ✓ | | ✓ | | | |
| Kuroda et al. [60] | 2022 | ✓ | | | | | ✓ | ✓ | | | ✓ |
| Ustok et al. [120] | 2022 | ✓ | | ✓ | | | ✓ | ✓ | | | |
| Rivera et al. [93] | 2022 | | | ✓ | ✓ | ✓ | | ✓ | | | ✓ |
| Banerjee et al. [9] | 2021 | ✓ | | | | ✓ | | ✓ | | | |
| Christou et al. [22] | 2022 | ✓ | | | | | ✓ | ✓ | | | |
| Borsatti et al. [14] | 2022 | ✓ | | | | ✓ | | ✓ | | | |
| Baktir et al. [8] | 2022 | ✓ | ✓ | ✓ | ✓ | ✓ | | ✓ | | | |
| Saha et al. [94] | 2022 | ✓ | ✓ | | | ✓ | | ✓ | | | ✓ |
| Barrachina-Muñoz et al. [10] | 2022 | ✓ | ✓ | ✓ | ✓ | ✓ | | ✓ | | | |
| Zhang et al. [136] | 2022 | ✓ | | | | ✓ | | ✓ | | | |
| Xie et al. [130] | 2022 | ✓ | | | | ✓ | | ✓ | | | |
| Li et al. [63] | 2022 | ✓ | ✓ | | | ✓ | | ✓ | | | |
| Mi et al. [77] | 2022 | ✓ | | | | ✓ | | ✓ | | | |
| Zhang et al. [135] | 2022 | ✓ | ✓ | | | ✓ | | ✓ | | | ✓ |
| Chang et al. [18] | 2022 | ✓ | ✓ | | | ✓ | | ✓ | | | ✓ |
| Xiao et al. [129] | 2022 | ✓ | | | | ✓ | | ✓ | | | |
| Mehmood et al. [74] | 2022 | ✓ | | | | ✓ | | ✓ | | | |
| Angi et al. [7] | 2022 | ✓ | | | | ✓ | | ✓ | | | ✓ |
| Souihi et al. [108] | 2022 | ✓ | | | | ✓ | | ✓ | | | ✓ |
| Alcock et al. [5] | 2022 | | | ✓ | ✓ | ✓ | | ✓ | | | |
| Meijer et al. [75] | 2022 | ✓ | | | | ✓ | | ✓ | | | ✓ |
| Teng et al. [113] | 2022 | ✓ | | | | ✓ | | ✓ | | | ✓ |
| Song et al. [107] | 2022 | ✓ | | | | ✓ | | ✓ | | | |
| Chowdhary et al. [21] | 2022 | ✓ | | | | | ✓ | ✓ | | | |
| Karrakchou et al. [51] | 2022 | ✓ | | | | ✓ | | ✓ | | | |
| Lin et al. [64] | 2022 | ✓ | | | | | ✓ | ✓ | | | |
| Ribeiro et al. [90] | 2022 | ✓ | | | | ✓ | | ✓ | | | |
| Hireche et al. [45] | 2022 | ✓ | | ✓ | ✓ | | ✓ | ✓ | | | ✓ |
| Zheng et al. [138] | 2022 | ✓ | | ✓ | ✓ | ✓ | | ✓ | | | |
| Martini et al. [68] | 2022 | ✓ | | ✓ | ✓ | | ✓ | ✓ | | | |
| Ooi et al. [84] | 2022 | ✓ | | | | ✓ | | ✓ | | | |
| Wu et al. [126] | 2022 | ✓ | ✓ | ✓ | | ✓ | | | ✓ | | ✓ |
| Sharma et al. [100] | 2022 | ✓ | | | | | ✓ | ✓ | | | |

**A1:** Activity 1 (Intent specification and translation), **A2:** Activity 2 (Autonomous deployment and orchestration), **A3:** Activity 3 (Monitoring and awareness)
**A4:** Activity 4 (Dynamic optimization and remediation), **SD:** Single domain, **MD:** Multi Domain, **ML:** Machine Learning

*6.1.1 Lack of Complete Solution.* Research in IDSM systems is at early stage; there is a lack of a comprehensive solution covering all four activities of intent management (IDSM Activities section of Table 10). Figure 19.1 shows the activity wise distribution of the research works considered in this study. Given figure makes it clear that research in IDSM systems has concentrated primarily on Intent Specification and Translation (Activity 1). 62 out of 104 works (58%) covers only Activity 1 in their proposed IDSM solutions. Only three complete solutions (CS) proposed by Yang et al. [131], Baktir et al. [8] and Barrachina-Muñoz et al. [10] covers all the four activities.

*6.1.2 Intent Management in Multiple Domains/Sub-systems.* Adoption of technologies, such as intent-driven interfaces, closed loop automation and knowledge driven decision making (based on AI and ML), increases the complexity of IDSM systems. To reduce such complexity, IDSM systems can be arranged into layers separating business, service and resource operations; and deployed in multiple domains/subsystems that can operate autonomously. All the layers and domains/sub-systems work together in a closed loop manner and, interact and coordinate with each other by using intent handlers (IHs) to fulfill the intents (Figure 3). However, it has been observed that majority of solutions do not consider the multi-layer and multi-domain architecture and focused only on the single domain/sub-system solutions (Figure 19.2). In the proposed multi-domain solutions (Table 10), the interaction and intercommunication between the IHs of different layers and domains/sub-systems either remained untouched or partly explored.

*6.1.3 Network-Centric Solutions.* As discussed in the background section (Section 2), apart from the networking field, the adoption of IDSM has been explored in other fields, such as cloud computing. However, from the current state of the art (Table 10), it has been observed that the majority of the intent-driven solutions are focused on the network service management. Very few solutions are available for other fields, such as cloud computing and block-chain (Figure 19.3). Not having intent-driven solutions for such critically important fields will hinder the development of complete IDSM solutions and could limit the value and adoption of the technology.

*6.1.4 Use of Machine Learning.* ML is becoming ubiquitous owing to the availability of massive data and improvement in computing power and algorithm innovation. Because of this, ML plays an important role in many fields, including computing and network operations. Considering the hierarchical and multi-domain characteristics of IDSM systems, integration of 'operational intelligence' by using ML methods at each layer (business, service and resources) to achieve closed loop autonomy is an ultimate goal [83]. Despite this, the current state-of-the-art for IDSM systems has the limited use of ML methods and relies heavily on static solutions for intent management (Table 10). However, an increasing trend of employing the ML methods for intent management is seen in the research articles published from 2020 to 2022, which accounts 78% of the total solutions using ML (Figure 19.4).

## 6.2 Open Challenges and Future Directions

We have identified various challenges which can be used to drive the future research in the area.

*6.2.1 Intent Negotiation Framework:* An intent submitted by a user may conflict with the service provider intent or with the intents submitted by other users. To resolve the conflicts, intent negotiation (Figure 10) takes place either between the human user and intent handler (IH) or among the IHs (either at the same level or different levels in the hierarchy). During intent negotiation, alternate intents representing the current capability of the service provider are generated and provided to the user or IH to select from. In order to do so, an intent negotiation framework is required providing a procedure to extract the state of the system and to use it to compose the alternate intents.

*6.2.2 Decomposition of Non-Functional Attributes:* Decomposition of functional attributes of an intent (for example, selection and chaining of VNFs to satisfy an intent) can be performed by using a knowledge-base consisting of ontologies. However, the decomposition of non-functional attributes and distribute them between the entities obtained after the functional decomposition is a cumbersome process. As shown in Figure 11, set of VNFs/PNFs and their deployment domains/sub-systems (edge, communication service provider (CSP) and cloud) are identified to satisfy the intent during functional decomposition. Now, the challenge is to decompose the quantitative values of non-functional attributes (latency, cost and availability) between edge, CSP and cloud sub-systems (Figures 12-15). This should be done while meeting the service level agreement (SLA) requirements of the original intent. A mechanism is required to perform an efficient decomposition of non-functional attributes of an intent without impacting the aggregated requirements of the original intent.

*6.2.3 Comparison of System KPIs and Intent's Non-Functional Requirements:* For SLA compliance, it is required to collect the relevant system KPIs and aggregate them to get the values corresponding to the non-functional attributes of an intent. Then, these values are compared with the expected values in the original intent to measure the its satisfaction level. A method is required to carry out such collection, aggregation and comparison operations optimally with minimal processing overheads to measure the real-time quality of experience (QoE) of the intent owners.

*6.2.4 Inter-operations between Legacy and Intent-driven Systems:* With the advancements of intent-driven service management (IDSM) systems, more service providers will start switching from the traditional methods of service providing and management to intent-driven methods. However, it won't be possible to perform such transition in one go and will happen in a progressive manner. To support such transition period, mechanisms are required to enable the inter-operations between the legacy and IDSM systems. The mechanisms involve the development of integration adapters able to map the requests between both kind of the systems.

*6.2.5 Standardized and Generic Intent Specification Method:* A standard method and template is required for the intent specification [82]. This will help to remove the current multi-vendor differences, such that all the existing IDSM solutions have their own intent specification methods. This does not allow the inter-working of these solutions and make them platform dependent, which results in vendor lock-ins. Having a standard and generic intent specification template can simplify the integration of multi-vendor systems required to enable the service.

## 7 CONCLUSIONS

The concept of Intent-driven service management (IDSM) has recently been proposed with a goal to simplify the deployment and management of network and computing services. This is achieved by transiting from traditional human-driven service management to zero-touch service management. In IDSM, service level agreement (SLA) requirements are specified in a declarative manner as 'intents' which are then fulfilled, autonomously by using closed control-loop operations. As a result, the errors and misconfigurations caused by human-driven manual operations reduce significantly, making service deployments faster, cheaper and improves the quality of service (QoS). However, the IDSM systems are still in their beginning phase. Hence, there is a need to identify and develop a deep understanding of what are their main components and which activities they performed to manage and fulfill the intents? While answering these questions, we reviewed the existing methods and solutions proposed for IDSM systems. As a result, we proposed a conceptual multi-layered and multi-domain architecture for IDSM systems. Additionally, we identified four activities the IDSM systems perform to fulfill the intents. For each activity, separate taxonomies are proposed.

Existing SLA management solutions for IDSM systems are compared and investigated based on these taxonomies. This allowed us to identify the research gaps in the state-of-the-art and propose various future research directions. As a result, we assert the following conclusions:

- IDSM systems perform four activities to fulfill the intents: intent specification and translation, autonomous deployment and orchestration, monitoring and awareness, and dynamic optimization and remediation.
- Developing a generic IDSM framework to represent intent processing from its specification to its fulfillment is necessary to manage the SLAs effectively. This will standardize the intent processing operations and their interplay.
- To accommodate the diversified needs of the service users and their SLAs, multi-vendor and multi-domain IDSM solutions should be developed by intensifying the interface standardization and development of integration adaptors.